\documentclass[%
 reprint,
 amsmath,amssymb,
 aps,
]{revtex4-2}

\usepackage{graphicx}
\usepackage{dcolumn}
\usepackage{bm}

\DeclareMathOperator{\sech}{sech}

\usepackage{xcolor}

\begin{document}

\preprint{APS/123-QED}

\title{Transitions between polarisation and radicalisation in  a temporal bi-layer echo chambers model}

\author{\L{}ukasz G. Gajewski}
 \email{lukaszgajewski@tuta.io}
\affiliation{Faculty of Physics, Warsaw University of Technology, Koszykowa 75, 00-662 Warszawa, Poland}
 \author{Julian Sienkiewicz}
 \affiliation{Faculty of Physics, Warsaw University of Technology, Koszykowa 75, 00-662 Warszawa, Poland}
 \author{Janusz A. Ho\l{}yst}
 \affiliation{Faculty of Physics, Warsaw University of Technology, Koszykowa 75, 00-662 Warszawa, Poland}
 \affiliation{ITMO University, Kronverkskiy Prospekt 49, St Petersburg, Russia 197101}






\begin{abstract}
Echo chambers and polarisation dynamics are as of late a very prominent topic in scientific communities around the world. 
As these phenomena directly affect our lives and seemingly more and more as our societies and communication channels evolve it becomes ever so important for us to understand the intricacies of opinion dynamics in the modern era. 
Here we extend an existing echo chambers model with activity driven agents onto a bi-layer topology and study the dynamics of  the polarised state as a function of interlayer couplings. Different cases of such  couplings are presented - unidirectional coupling that can be reduced to a mono-layer facing an external bias, symmetric and non-symmetric couplings.
We have assumed that  initial conditions impose system polarisation  and agent opinions are different for both layers. 
Such a pre-conditioned  polarised state can sustain   without explicit homophilic interactions provided the coupling strength between agents belonging to different layers is weak enough. For  a strong unidirectional or attractive  coupling between two layers a discontinuous transition to a radicalised state takes place when mean opinions in both layers are the same.    
When coupling constants between the layers are of different signs the system exhibits sustained or decaying oscillations. Transitions between these states are analysed using a mean field approximation and  classified in the framework  of bifurcation theory.

\end{abstract}

\maketitle


\section{\label{sec:level1}Introduction\protect}
It is not unheard of in the scientific community to attempt to model how our societies form and function using techniques and approaches familiar to physicists \cite{holyst2001social, castellano2009statistical, serge2016sociophysics, Galam2013ModelingTF, sen2014sociophysics}. 
Of a particular interest lately has been the dynamics of opinion formation, especially in the light of recently better studied phenomena such as echo chambers \cite{Baumann_2020, cota2019quantifying, tornberg2018, Sasahara_2020} and misinformation \cite{delVicario_pnas, Vosoughi1146, bessi_trend_2015, shao_anatomy_2018, shao_hoaxy_2016, shao_spread_2018}. 
One of the major effects that seems to be strongly connected with echo chambers and misinformation is that of polarisation. While not every topic is polarising \cite{dimaggio1996have, fiorina} many certainly can be \cite{mouw2001culture, mccright, lin, cota2019quantifying, conover2011political, conover2012partisan, hanna, brady2017emotion, weber2013secular}. 
It seems to have been recognised by some as dangerous to the state of democracy around the world, and that there is a need for research in this topic \cite{baumann2021emergence, schweighofer2020, math7100917, kurahashi2016robust, banisch2019opinion, Macyeaax0754, gorski}. Especially in the light of a possible event of democracy backsliding \cite{Waldner2018,Wiesner2018}.

We find that it is also of interest to study the possible dynamics between two clearly defined groups as it often can be in politics (e.g. Democrat vs Republican in the USA), topics (pro- or anti-) as well as has precedence in sociophysics \cite{Suchecki_2006, suchecki2009,krueger, chmiel, choi2019majority, lambiotte2007coexistence, lambiotte2007majority}. In particular we felt inspired by the work of Baumann et al. \cite{Baumann_2020} where the authors introduce an echo chambers and polarisation model on complex networks. In this paper we modify said model so that it operates on a bi-layer temporal network, as opposed to a mono-layer, where each layer can represent a clearly defined group of individuals (agents). This transformation is directly driven by the fact that many system drastically change their physical properties (e.g., phase transition type change) when considered on a duplex (bi-layer) topology \cite{Chmiel2017,Chmiel2020}. We show that several complex behaviours can be acquired by simply changing the nature of the coupling between those layers. Let us underline that the question of interacting layers is an extremely vivid topic in the view of COVID-19 epidemic (or infodemic \cite{Gradon2021}). Recent studies point to a pivotal role played by risk perception layer in the spreading of a disease \cite{Ye2020} or explicitly the attitude toward vaccination \cite{zuzek2017}. In this scope examining the dynamics of two {\it coupled} opposite groups (e.g., pro- and anti-vaccination \cite{Johnson2020}) seems to be highly relevant.   

Originally, in the work of Baumann et al., the system consists of $N$ agents each with a real, continuous opinion variable $x_i(t) \in \mathbb{R}$. The sign determines the nature of opinion (for/against) while the value the conviction to it. The opinion dynamics is driven exclusively by the interactions between agents and is described by a system of coupled ordinary differential equations presented in \cite{Baumann_2020}:
\begin{equation}
    \Dot{x}_i = -x_i + K \sum_{j=1}^{N}A_{ij}(t)\tanh{(\alpha x_j)},
    \label{eq:dynamics}
\end{equation}
where $K > 0$ is the \textit{social interaction strength} and $\alpha > 0$ determines the degree of non-linearity.
The rationale behind this very equation is built on the mechanism of informational influence theory with guarantees of monotonic influence and a cap on extreme opinions while also not being dissimilar to previously used non-linear functions in chaotic systems \cite{mas, Jayles12620, sompolinksy, eldar2013effects}.

\begin{figure*}[htb]
         \includegraphics[width=0.95\textwidth]{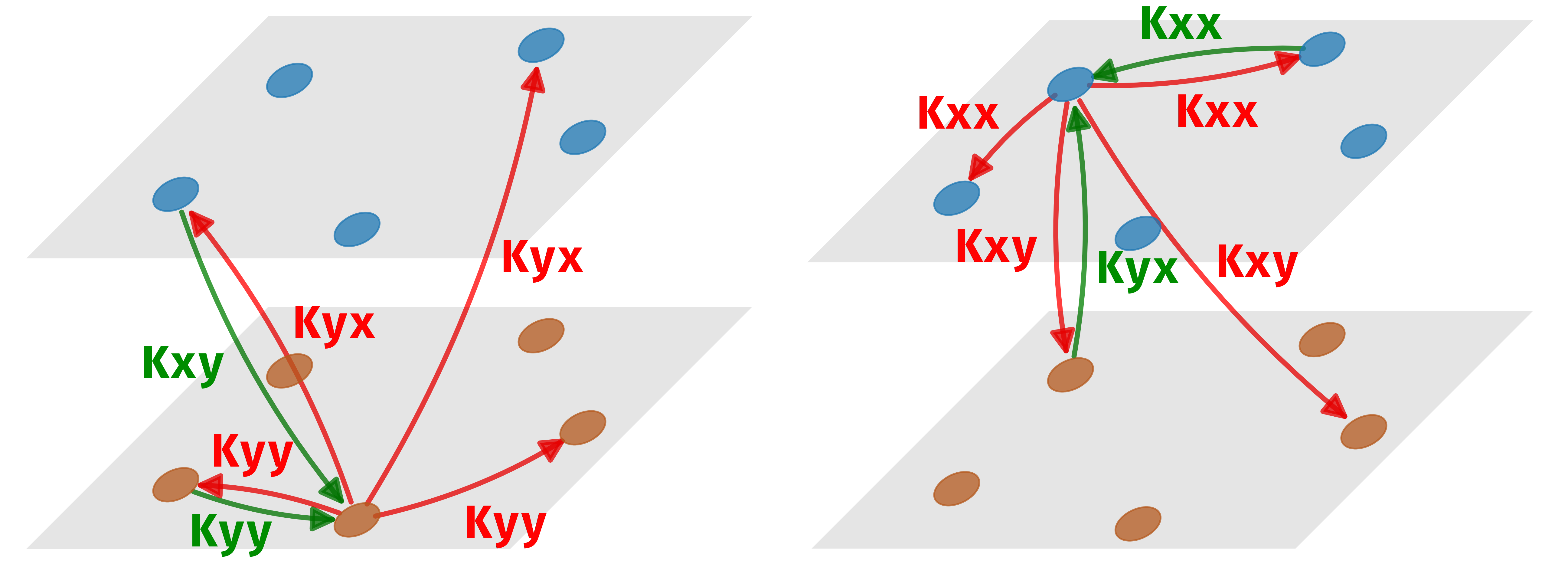}
\caption{Illustration of the temporal bi-layer network model. At any given moment an agent from either group can get activated and impose its influence upon other (red arrows) while in some cases this influence can be reciprocated (green arrows). Each arrow is labelled with an appropriate social influence coefficient later on used in system of equations~(\ref{eq:2groups}).}
\label{fig:arrow}
\end{figure*}
The matrix $A$ is an $N\times N$ adjacency matrix in an activity-driven (AD) temporal network model \cite{perra2012activity, starnini, moinet, liu} (see Fig.~\ref{fig:arrow} for our bi-layer interpretation).
This is a model without a statically set social network but in each time step an agent can become active with \textit{propensity} $a_i \in [\epsilon, 1]$. The propensities are drawn from a power law distribution \cite{perra2012activity, moinet} defined as follows:
\begin{equation}
    F(a) = \frac{1-\gamma}{1-\epsilon^{1-\gamma}}a^{-\gamma}.
    \label{eq:activity}
\end{equation}
Once the agent is activated it makes $m$ random connections with other agents and as is standard in AD models the connections are uniformly random.
In \cite{Baumann_2020} there is additionally an element of homophily as it is expected to be necessary to create polarisation effects \cite{holme, kimura}, however, since we will be considering a bi-layer model later on this is not the case for us. A proper study of the effects of the homophily in the from presented by Baumann et al. could turn out to be of interest yet we find it going beyond the scope of this paper.

The interactions in social media can often be asymmetric and so it is not always true that $A_{ij} = A_{ji}$. However, in this model there is a mechanism of \textit{reciprocity} where each agent $j$ that has received a connection from an active agent $i$ can reciprocate the connection with probability $r$.

Following the terminology from the paper \cite{Baumann_2020} we will call three specific opinion distributions as follows: (i) {\it a neutral  consensus} (or simply consensus)  will correspond to a phase when agents' opinions at both layers are similar and in average are close to zero; (ii) {\it a one-side  radicalisation} OSR (or simply radicalisation) will be  the opinion distribution when at both layers either a positive or a negative opinion is overwhelming and  it is the same at both layers;      (iii) {\it a polarisation} will be  the opinion distribution when at one layer a positive opinion is overwhelming but on the other layer the negative opinion  is overwhelming. Using the language of magnetic systems the neutral consensus corresponds to the paramagnetic phase, the radicalisation is the ferromagnetic phase and the polarisation is the antiferromagnetic phase \cite{mattis2012theory}.

\section{Model description}
We modify the scenario described by Baumann et al. by considering a system of two (potentially) opposing groups represented by layers - X and Y - such that $N_{X}$ of agents belong to group X and $N_{Y}$ to Y.
With this (\ref{eq:dynamics}) becomes:
\begin{widetext}
\begin{equation}
\begin{cases}
    \Dot{x}_i = -x_i + K_{xx} \sum\limits_{j}^{N_{X}}A_{ij}^{xx}(t) \tanh(\alpha_{xx} x_{j}) + K_{xy} \sum\limits_{j}^{N_{Y}} A_{ij}^{xy}(t)\tanh(\alpha_{xy} y_{j}) \\
    \Dot{y}_i = -y_i + K_{yy} \sum\limits_{j}^{N_{Y}}A_{ij}^{yy}(t) \tanh(\alpha_{yy} y_{j}) + K_{yx} \sum\limits_{j}^{N_{X}} A_{ij}^{yx}(t)\tanh(\alpha_{yx} x_{j}),
\end{cases}    
    \label{eq:2groups}
\end{equation}
\end{widetext}
This is the most general formulation of the model we propose and we will now appropriately simplify it as well as later on discuss its various regimes and scenarios that emerge from it.

Let us further assume $K_{xx}=K_{yy}=K$, $\alpha_{xx} = \alpha_{yy} = \alpha_{xy} = \alpha_{yx} = \alpha$, $N_{X} = N_{Y} = \frac{1}{2}N$ and both $r$ and $a$ to be the same for both groups, within as well as without.
Average activity is given by:
\begin{equation}
    \langle a \rangle = \frac{1-\gamma}{2-\gamma}\frac{1-\epsilon^{2-\gamma}}{1-\epsilon^{1-\gamma}}.
\end{equation}

Similarly as in \cite{Baumann_2020} we assume that processes related  to topology changes as described by matrices $A_{ij}(t)$ are much faster than changes of opinions $x_i(t)$ and $y_i(t)$ and we shall insert into (\ref{eq:2groups}) mean values of these matrices  $\langle A_{ij}(t) \rangle_{t,a} = \frac{1}{2} m (1+r) \langle a \rangle$, see Section ``Approximation of the critical controversialness'' in the supplemental materials of \cite{Baumann_2020} for detailed derivation.    
When $K_{xy}=K_{yx}$ then the Jacobian of (\ref{eq:2groups}) calculated in the point $x_i=y_i=0$ possesses  two special eigenvectors, $\boldmath{e}_+=[1,1,1...1,1,1]^T$ and  $\boldmath{e}_-=[1,1,1...-1,-1,-1]^T$ and corresponding eigenvalues $\lambda_+=c\alpha[K(N_x-1)/N_x +K_{xy}]$ and $\lambda_-=c\alpha[K(N_x-1)/N_x -K_{xy}]$.

Then we can write mean field equations for the expected values of opinions in $X$ and $Y$.\\
For simplicity let us set $c = \frac{m}{2}(1+r)\langle a \rangle$ and then,
\begin{equation}
    \begin{cases}
    \Dot{\langle x \rangle} = - \langle x \rangle + K c \tanh(\alpha \langle x \rangle) + K_{xy} c \tanh(\alpha \langle y \rangle)\\
    \Dot{\langle y \rangle} = - \langle y \rangle + K c \tanh(\alpha \langle y \rangle) + K_{yx} c \tanh(\alpha \langle x \rangle).
    \end{cases}
\label{eq:system}
\end{equation}

We show that in our bi-layer variant of the echo chambers and polarisation model \cite{Baumann_2020}, when initial conditions impose a polarised state and there are opposite  agents' opinions at different layers then depending on the type of inter-layer coupling various  patterns are observed.  For a weak attractive  coupling the polarised state is preserved but when the coupling reaches a critical value a discontinuous transition to a radicalisation phase  \cite{Baumann_2020,myers1976group,isenberg1986group} takes place and opinions  at both layers are similar and biased  towards a positive or negative value. An asymmetric  (attractive/repulsive) coupling between agents at  both layers induces oscillations of  opinions.    


Let us stress that when the coupling between layers is weak then  the layers  operate independently and  each of them  becomes an analog to the system studied in \cite{Baumann_2020} when the homophily is neglected.  Thus, each given layer can be  radicalised but the  composite bi-layer system can be also  {\it polarised} when each layer has its own radicalised state {\it opposite} to the other layer’s (this opposition state depends on the initial conditions, however).

Further on we provide agent-based simulations and detailed mathematical analysis that makes use of  the mean field approximation and catastrophes theory  and well fits to results of  agent-based  numerical simulations.

Our work is also distinctly different from the recent publication of Baumann et al.~\cite{baumann2021emergence} where authors consider a multidimensional version of the echo chambers model. In their work the coupling occurs via a correlated \textit{topic} space wheres we establish a variant with \textit{interacting groups}, quite naturally leading to very different phenomena being observed.
\section{Methodology}
All simulations were conducted, unless stated otherwise, with parameter values: network size $N=1000, \gamma=2.1, \epsilon=0.01, m=10, r=0.5, K=1, \alpha=1, K_{xy}=K_{yx}=-1$ (or 1, -1 accordingly in the asymmetric, oscillating case and 1,1 in the positive symmetric, weak coupling case). Note that as a consequence of these values the parameter $c \approx 0.306$.
The systems of equations in the agent-based simulations were integrated using an explicit fourth order Runge-Kutta method with a time step $dt = 0.05$. The temporal adjacency matrix $A_{ij}$ is computed at each integration step. Mean field equations where no analytical solution was possible were integrated using an embedded Runge-Kutta 5(4)\cite{DORMAND198019, 2020SciPy-NMeth}.
Following the rationale in \cite{Baumann_2020, krosnick1988attitude} the AD network is updated on each integration step as to separate the timescales of connections and opinion dynamics.

\section{Results}
In this section we present the results of agent-based simulations and the mean field approximation to the four scenarios described before. The scenarios are: (a) unidirectional coupling (this case will be  equivalent to an external bias), (b) symmetric coupling, (c) non-symmetric coupling.

\subsection{Unidirectional coupling}
We can study a cumulative effect of a bi-layer environment via an addition of external bias to a mono-layer system. This bias can represent cumulative effect of another group (Y) or just the medium in which the system operates. 

In essence, stemming from Eq.~(\ref{eq:2groups}), we set $K_{yx}=0$, $K_{xx} = K \neq K_{yy}$, $\alpha_{xx} = \alpha \neq \alpha_{yy} \neq \alpha_{yx} \neq \alpha_{xy}$. If $K_{yy}\alpha_{yy} c>1$ then the layer  Y is radicalised and agents' opinions $y_i$ in this layer are centered around a certain nonzero value $\langle y \rangle$ that is constant in time. In such a case  the whole term $K_{xy} \sum\limits_{j}^{N_{Y}} A_{ij}^{xy}(t)\tanh(\alpha_{xy} y_{j})=B_i$ can  be ``hidden'' behind a cumulative effect - an external bias $B_i$ 
 that can  be in general dependent on the site $i$ and can be   either supporting a local opinion $x_i$ in the layer X or working in opposition to $x_i$. 

\begin{figure*}[htb]
        \includegraphics[width=\textwidth]{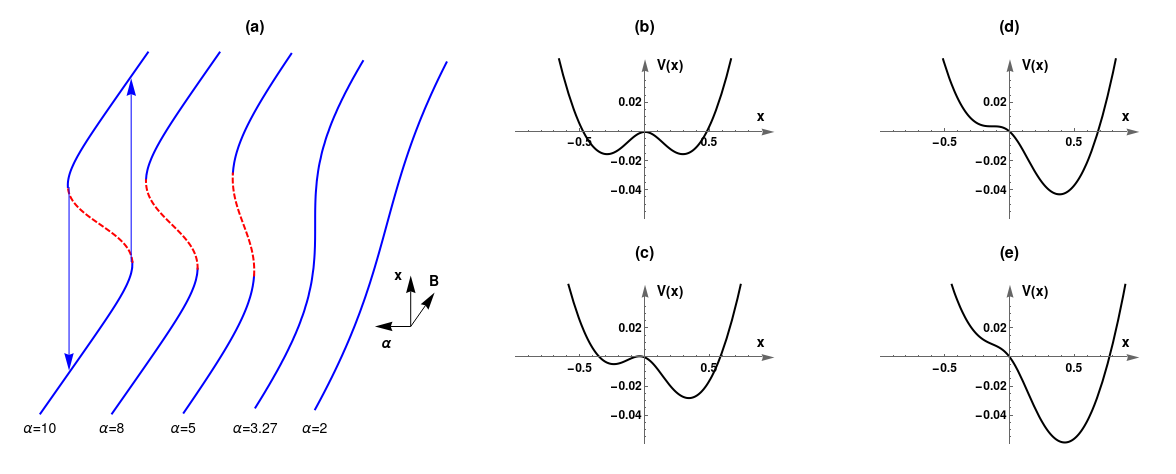}
\caption{(a) Bifurcation and hysteresis loop in the system of external bias: for $\alpha > 1/(K c)\approx 3.27$ the system can be  bistable and once a critical value of $B_c$ is reached there is a switch of opinion majority  from a state  against the field to towards it (an vice-versa for $-B_c$).  Also in such a case, we cannot reach a neutral solution ($x=0$) for any $B > 0$. For $\alpha < 1/(K c)$ we have only one stable solution and such effects do not take place. (b-e) Shape of the potential $V(x)$ given by Eq. (\ref{eq:vx}) for $Kc\alpha=2$ and $B=0$ (b), $B=0.04$ (c) $B=0.0815$ (d) and $B=0.12$ (e).}
\label{fig:bif}
\end{figure*}

Therefore we can write that:
\begin{equation}
    \Dot{x}_i = -x_i + K \sum\limits_{j}^{N}A_{ij}(t) \tanh(\alpha x_{j}) + B_i,
    \label{eq:ext-field}
\end{equation}
and by averaging $x_i$ we get   

\begin{equation}
    \Dot{\langle x \rangle} = - \langle x \rangle + K c \tanh(\alpha \langle x \rangle) +B.
    \label{eq:mean_ext_field}
\end{equation} 
where $B= \langle B_i \rangle$.
The dynamical system described by  (\ref{eq:mean_ext_field}) exhibits a cusp catastrophe \cite{zeeman1979catastrophe, arnol2003catastrophe}. If $Kc\alpha <1$ then there is only one steady state solution of  (\ref{eq:mean_ext_field}). However if $Kc\alpha >1$ then two scenarios are possible.  
When the modulus of the external bias $B$ is smaller than some critical value $B_c$ then the  equation (\ref{eq:mean_ext_field}) possesses two stable and one unstable fixed point.  It means the mean opinion  in  the layer $X$  is in agreement or in disagreement  with the  external bias $B$.    
When $B$ is larger than some critical $B_c$ then the  equation (\ref{eq:mean_ext_field}) possesses only one  solution  and  the mean opinion in the group X directed against the external  bias B  is not possible. It means that at some critical  $B_c$  a discontinuous transition takes place (see Fig. \ref{fig:bif}a).  Values of $B_c$  can be found from the stability analysis of (\ref{eq:ext-field}) or (\ref{eq:mean_ext_field}).

In the case of  (\ref{eq:mean_ext_field}) we get the Lyapunov exponent \cite{lyapunov1992general} at the point   $x_c$ corresponding to a steady state solution that stability is examined      
\begin{equation}
    \lambda=   -1 + Kc\alpha~\sech^2(\alpha x_c).
    \label{eq:lambda}
\end{equation}

In the case of (\ref{eq:ext-field})  the Jacobian becomes:
\begin{equation}
J|_{x_i=x_c} = 
\begin{bmatrix}
 -1 & Kc\alpha~\sech^2(\alpha x_c) & \dots \\
 Kc\alpha~\sech^2(\alpha x_c) & -1 & \dots \\
 \vdots
\end{bmatrix},
\end{equation}
with the largest eigenvalue 
\begin{equation}
 \lambda_{max}  =  -1 + \frac{N-1}{N}Kc\alpha~\sech^2(\alpha x_c).
 \label{eq:lambda_Jac}
\end{equation}
When $N\to \infty$ then solutions (\ref{eq:lambda}) and (\ref{eq:lambda_Jac}) coincide.  Combining  the condition for the steady state of (\ref{eq:mean_ext_field}) and the condition for changing the sign of  the eigenvalue $\lambda_{max}$  (\ref{eq:lambda_Jac}) we get a solution for the critical value of the external bias $B_c$:
\begin{equation}
    \begin{cases}
        x_c = \frac{1}{\alpha}\cosh^{-1}\bigg(\sqrt{\frac{N-1}{N}Kc\alpha}\bigg) \stackrel{N \rightarrow \infty}{\sim} \frac{1}{\alpha}\cosh^{-1}\big(\sqrt{Kc\alpha}\big)\\
        B_c = x_c - Kc~\tanh(\alpha x_c),
    \end{cases}
    \label{eq:ext-field-crit}
\end{equation}

In order to explore the behavior of Eq.~(\ref{eq:ext-field}) one can examine the effective potential
\begin{equation}
    V(x) = -\int_{-\infty}^{x} F(u)du,
\end{equation}
where $F(x)$ is the so-called effective force, being r.h.s. of Eq.~(\ref{eq:ext-field}). Thus, in our case
\begin{equation}
    V(x) = \frac{x^2}{2} - \frac{Kc}{\alpha}\ln\cosh(\alpha x) - Bx.
\label{eq:vx}
\end{equation}
If $B=0$ (Fig. \ref{fig:bif}b) then the potential $V(x)$ is a symmetric function possessing two minimum values and one maximum, corresponding to, respectively stable and unstable solutions as long as $\alpha > 1/(Kc)$ or one minimum at $x=0$ if this condition is not fulfilled. However, if $B \neq 0$ the potential becomes asymmetric (Fig. \ref{fig:bif}c) and for $B \ge B_c$ 
the second minimum is no longer observed (see Fig. \ref{fig:bif}d-e). Let us note that if $Kc\alpha \gg 1$ in Eq.~(\ref{eq:ext-field-crit}) then $B_c \rightarrow -Kc$.   

\begin{figure*}[htb]
     \includegraphics[width=0.41\textwidth]{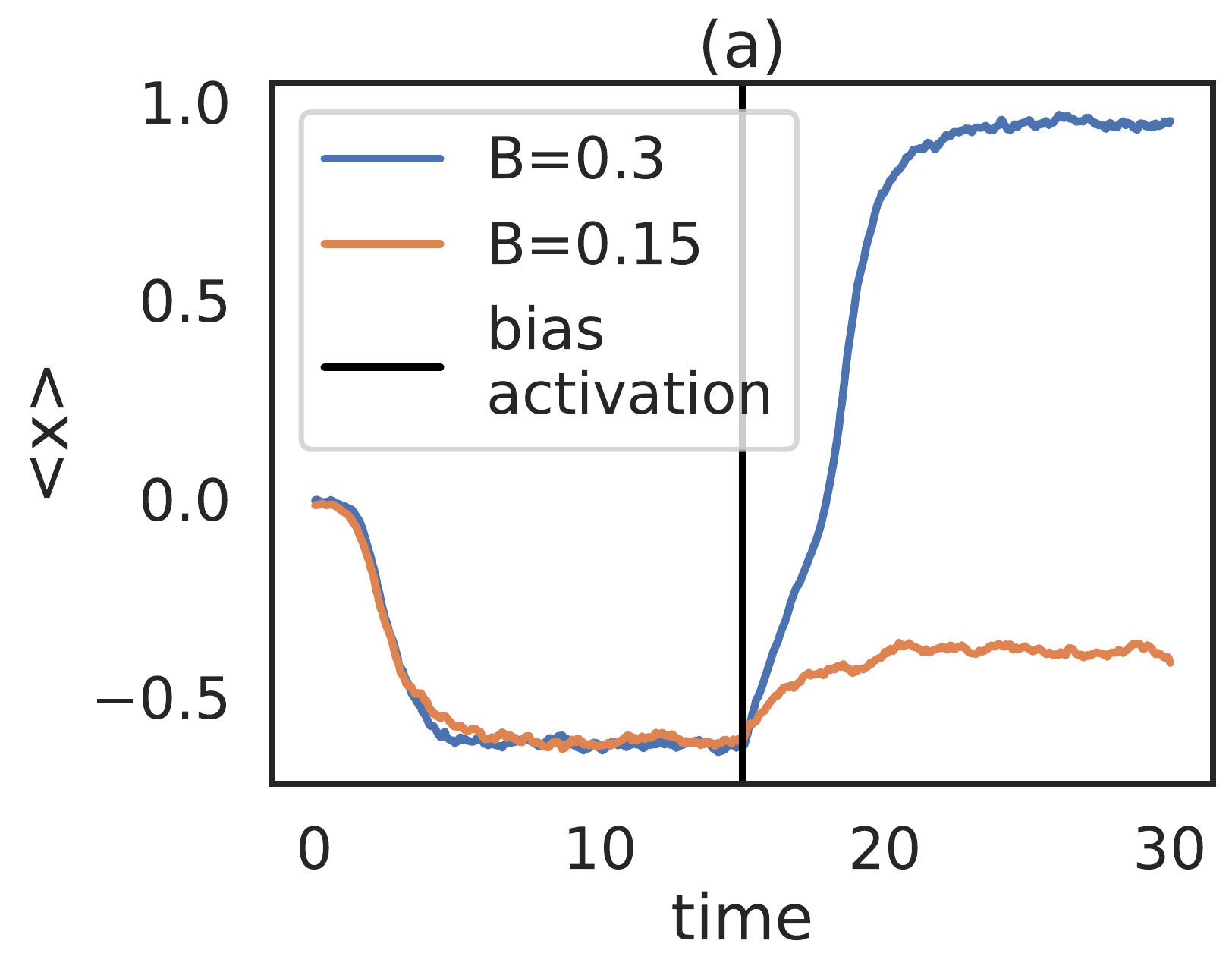}
     \hfill
    \includegraphics[width=0.43\textwidth, , height=0.25\textheight]{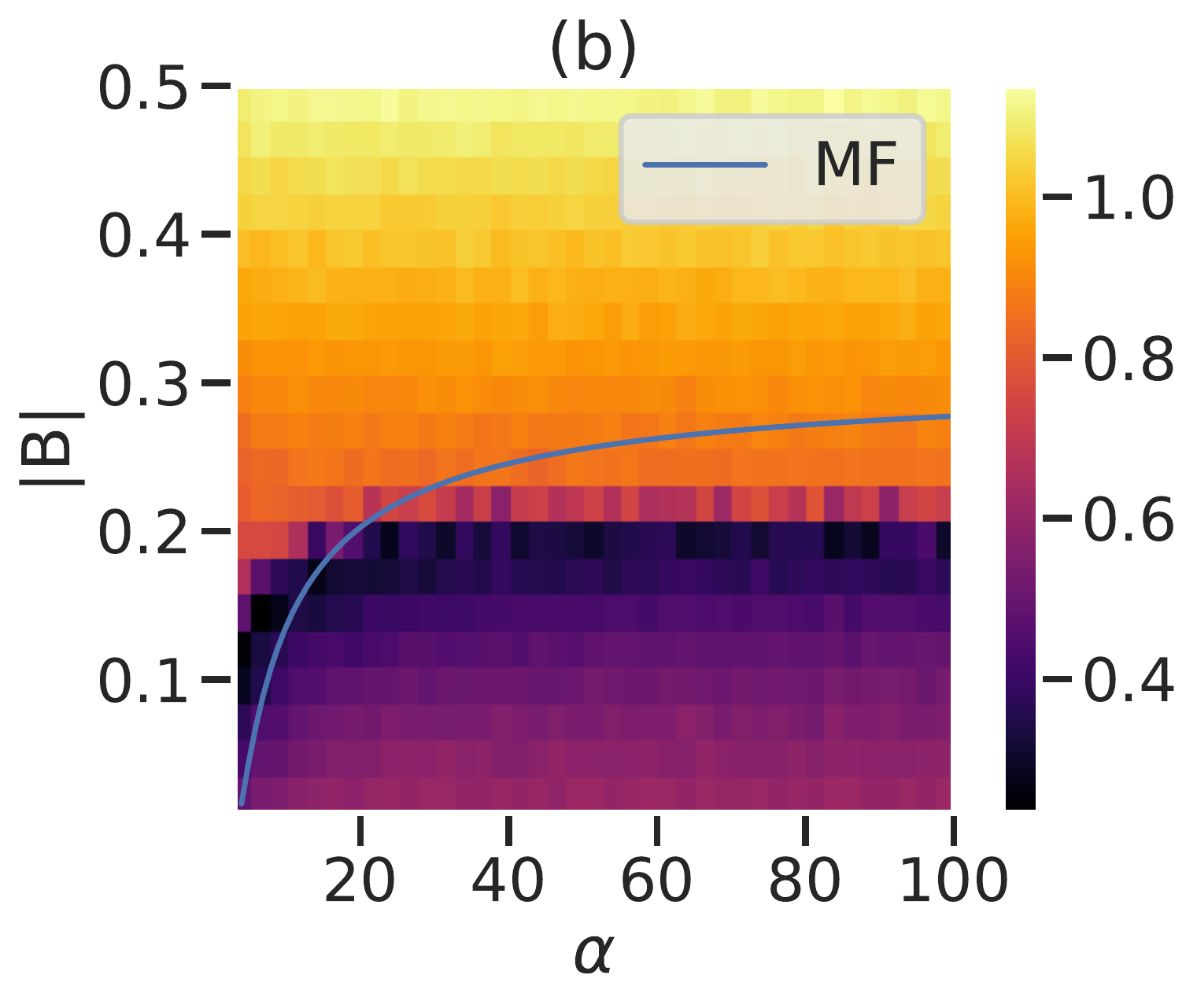}
\caption{Phase transition for the temporary network described by \ref{eq:ext-field} under the influence of an external bias. The left panel - (a) - shows two examples of an average opinion of the system as a function of time. One trajectory is for a value of external bias above the critical threshold and the other below. A solid vertical line signifying the moment we enable the external bias is present. The right panel - (b) - shows the $B-\alpha$ phase space, where colour is $\langle|$opinion$|\rangle$, with a visible phase transition to an opposite opinion and the mean field approximation for the critical line - Eq.~(\ref{eq:ext-field-crit}). }
\label{fig:ext_field}
\end{figure*}

The above results mean that a discontinous phase transition in the temporary network (\ref{eq:ext-field}) should occur from a  system's steady state to another one that is directed  towards the  external bias. E.g. if the system converges on a negative (average) opinion and we set the bias to a positive and sufficiently large value the system will suddenly jump to the opposite side. 
In Fig.~\ref{fig:ext_field}a we present an example of that. We wait until the system reaches its steady state and then activate the bias with an opposite sign. If the value is below the critical one the system merely shifts slightly towards zero, however, if $|B|>B_c$ a sudden jump occurs. 
In Fig.~\ref{fig:ext_field}b we show this in the $B-\alpha$ phase space: yet again the mean field approach --- Eq.~(\ref{eq:ext-field-crit}) --- allows us to predict this behaviour. 

We consider this case study as illustrative of how, for example, a propaganda may or may not be successful. We use ``propaganda'' here as a neutral term, without concerning ourselves whether it is good or bad. One can easily imagine situations that are either. Such a scenario boils down to the strength of the campaign in question since the dynamic of change is non-linear and the transition can be very sudden. One of the significant implications of this is that it may be rather difficult to react to the propaganda machine in time to stop the society from drastically shifting its stance.

\subsection{Symmetric coupling}
\label{sec:sym}
Here we consider a variant of the model when the two layers are positively but weakly coupled via the coupling parameter $\delta$. 
We introduce this weak-coupling parameter $0 < \delta < 1$ to the variant where both $K_{xy}$ and $K_{yx}$ are positive and for simplicity let us assume $K_{xy} = K_{yx} = \delta K$. 
Note that for large positive coupling $\delta > 1$ the system functionally reduces to the scenario already described by Baumann et al., and therefore will not be discussed by us. 
The mean field equations for the expected values can be written as:
\begin{equation}
    \begin{cases}
    \Dot{\langle x \rangle} = -\langle x \rangle + Kc~\tanh(\alpha \langle x \rangle) + \delta Kc~\tanh(\alpha \langle y \rangle) \\
    \Dot{\langle y \rangle} = -\langle y \rangle + Kc~\tanh(\alpha \langle y \rangle) + \delta Kc~\tanh(\alpha \langle x \rangle).
    \end{cases}
    \label{eq:delta}
\end{equation}

\begin{figure*}[htb]
         \includegraphics[width=0.48\textwidth]{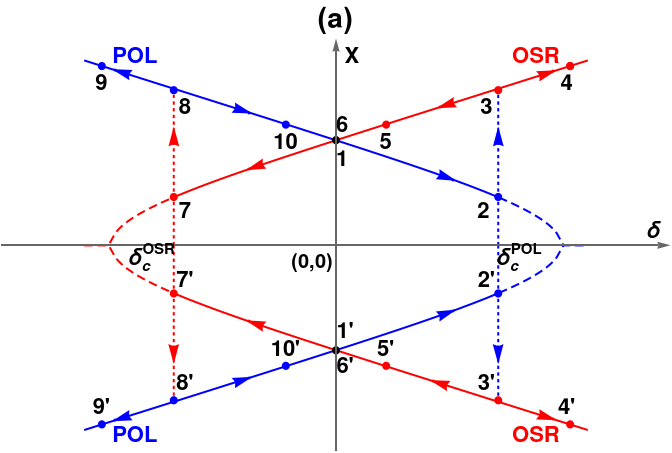}
         \hfill
         \includegraphics[width=0.48\textwidth]{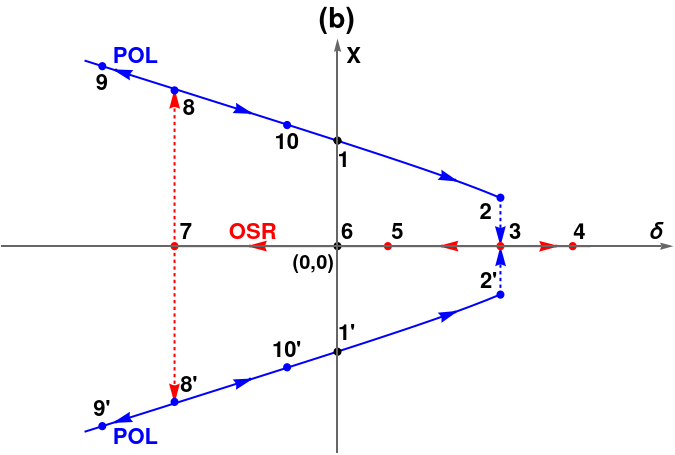}
         
     \includegraphics[width=0.45\textwidth]{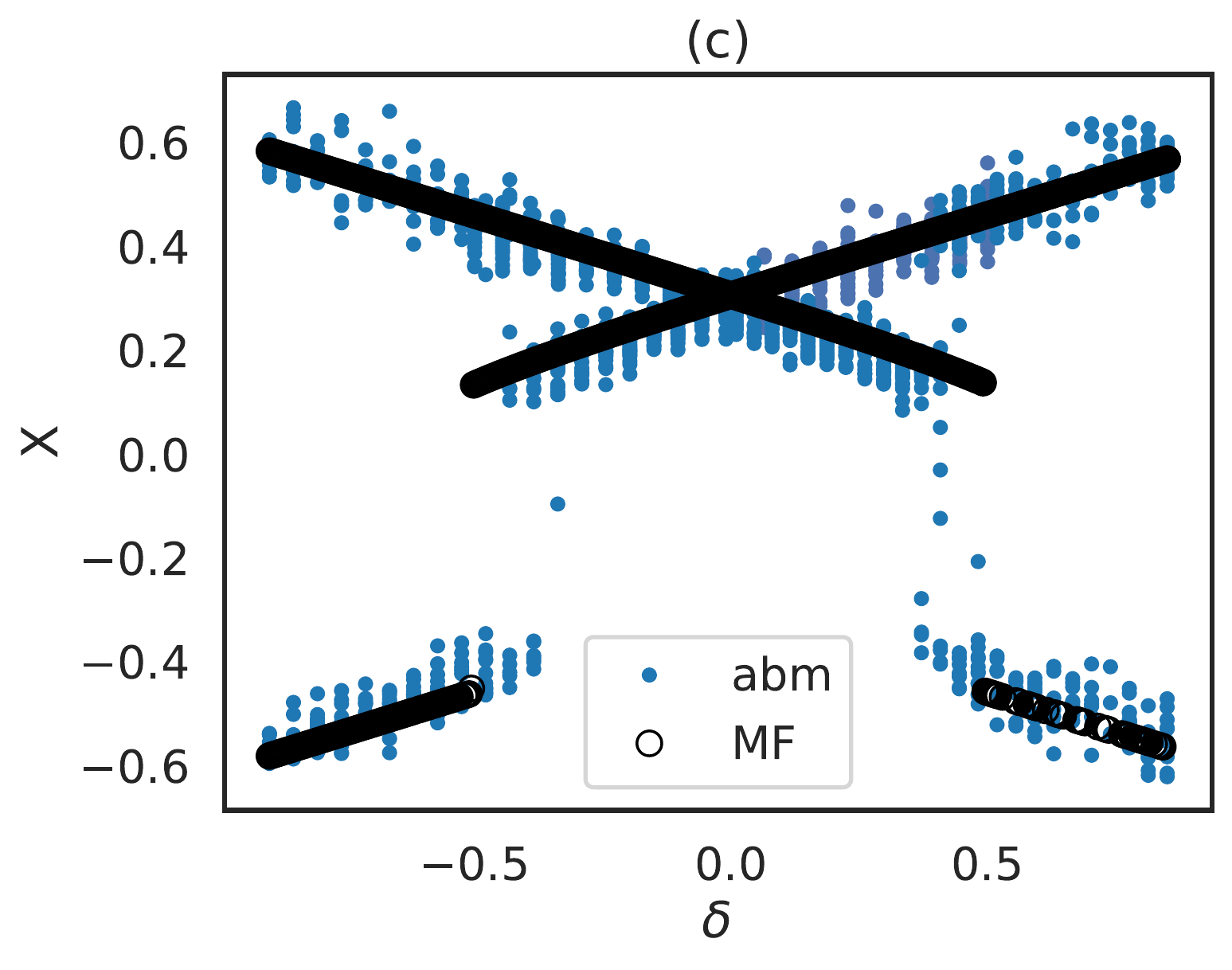}
     \hfill
         \includegraphics[width=0.43\textwidth]{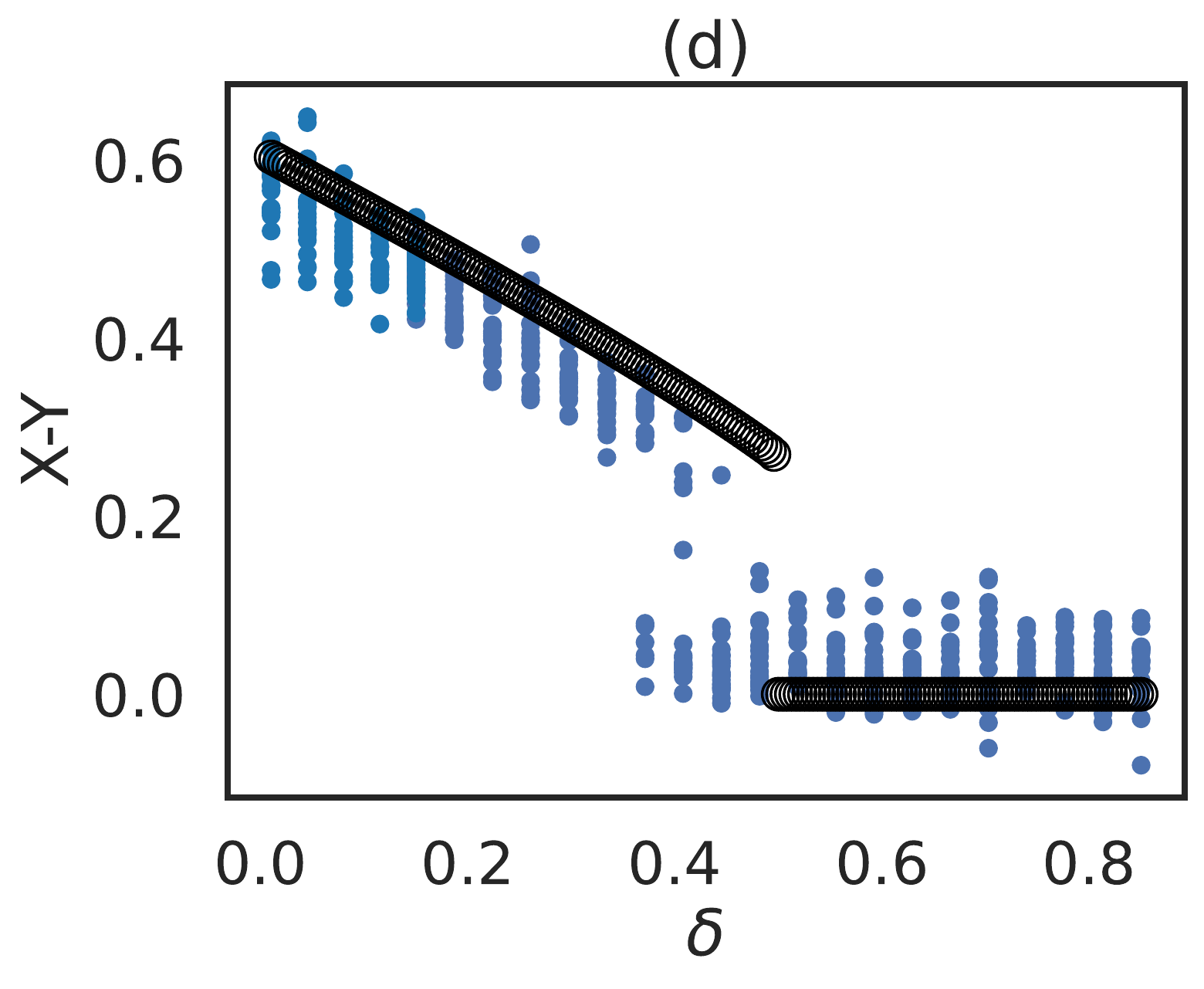}
\caption{(a-b) Illustration of the phase transition in a weakly coupled scenario for different values of $\delta$ in (a) $X,\delta$ and (b) $X-Y, \delta$ plane, see the main text for detailed description of points (1)-(10) and (1')-(10'); (c-d) Comparison of analytical predictions with ABM simulations. Dots show 20 independent realisations of an agent-based simulation while mean field solution is open circles (with $\alpha=10$). In (c) only the average opinion of layer X is shown (Y omitted for clarity as it would simply be symmetrically opposite).}
\label{fig:weakscheme}
\end{figure*}

\begin{figure*}[ht]
         \includegraphics[width=.77\textwidth]{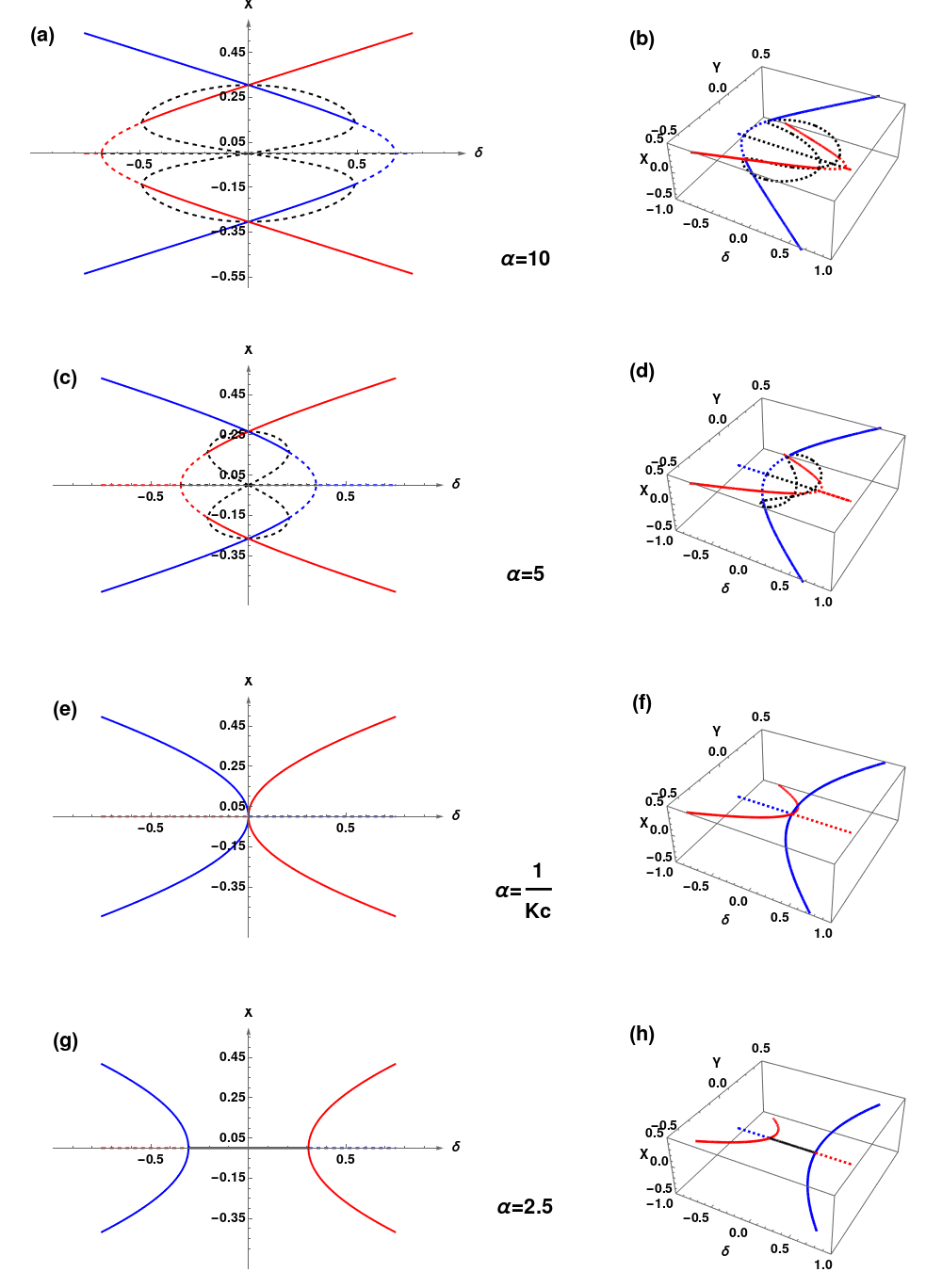}
\caption{Solutions of Eq.~(\ref{eq:2groups}) for different values of $\alpha$ (in each case $K=1$ and $c=0.306$) and : left column shows $x$ as a function of $\delta$ while the right one $x,y$ as a function of $\delta$. Solid lines represent stable solutions and dotted ones -- unstable. Red and blue curves denote radicalization and polarisation (as in Fig. \ref{fig:weakscheme}), black dashed lines show auxiliary solutions and black solid -- neutral consensus.} 
\label{fig:fig3D}
\end{figure*}

With positive coupling the two groups ought to merge for some critical value $\delta_c$. However, before that happens a coexistence of two groups with opposite opinions is possible. In such a case $x_c = -y_c$ in the steady state
and by writing out the Jacobian of the system (\ref{eq:delta}):
\begin{widetext}
\begin{equation}
J|_{\langle x \rangle=-\langle y \rangle=x_c} = 
\begin{bmatrix}
 -1 +  Kc\alpha\sech^2(\alpha x_c) & \delta Kc\alpha\sech^2(\alpha x_c) \\
 \delta Kc\alpha\sech^2(\alpha x_c) & -1 +  Kc\alpha\sech^2(\alpha x_c)
\end{bmatrix},
\label{eq:Jacob_sym}
\end{equation} 
\end{widetext}
from which we get both eigenvalues as:
\begin{equation}
    \lambda_{1,2} = Kc\alpha\sech^2(\alpha x_c) (1 \pm  \delta) - 1,
\end{equation}
and looking at the largest eigenvalue and the steady state solution it is easy to obtain that:
\begin{equation}
    \begin{cases}
    \delta_c = \frac{1}{Kc\alpha}\cosh^2(\alpha x_c) - 1 \\
    0 = -x_c + Kc(1-\delta_c)\tanh(\alpha x_c),
    \end{cases}
    \label{eq:delta-crit}
\end{equation}
which must be solved numerically.

We find that there exists a critical value $\delta_c$ for which a phase transition occurs from a polarisation (denoted as POL) state to a non-neutral consensus state (or the so called one side radicalisation -- OSR). 

\begin{figure*}[htb]
        \includegraphics[width=0.45\textwidth]{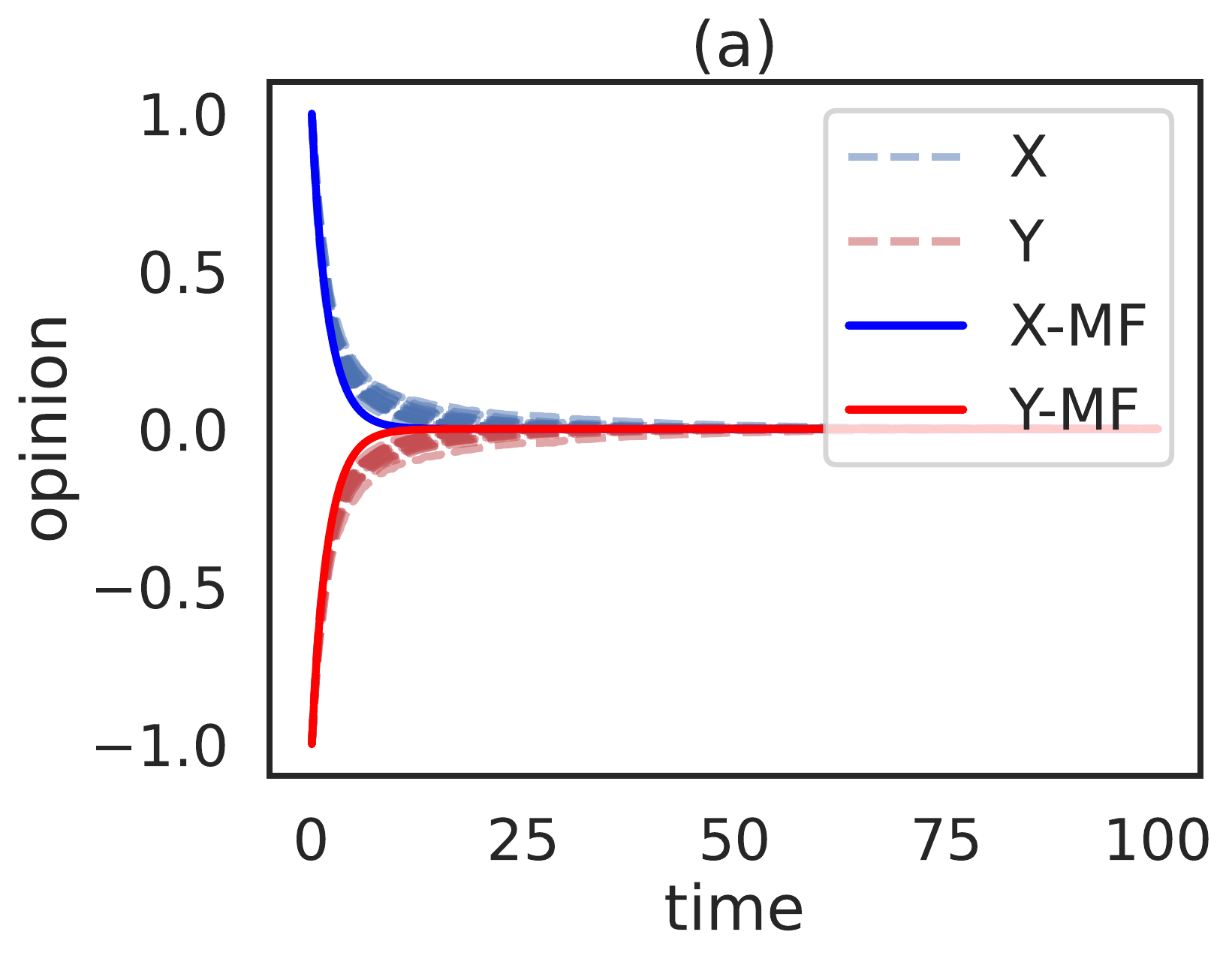}
        \hfill
        \includegraphics[width=0.45\textwidth]{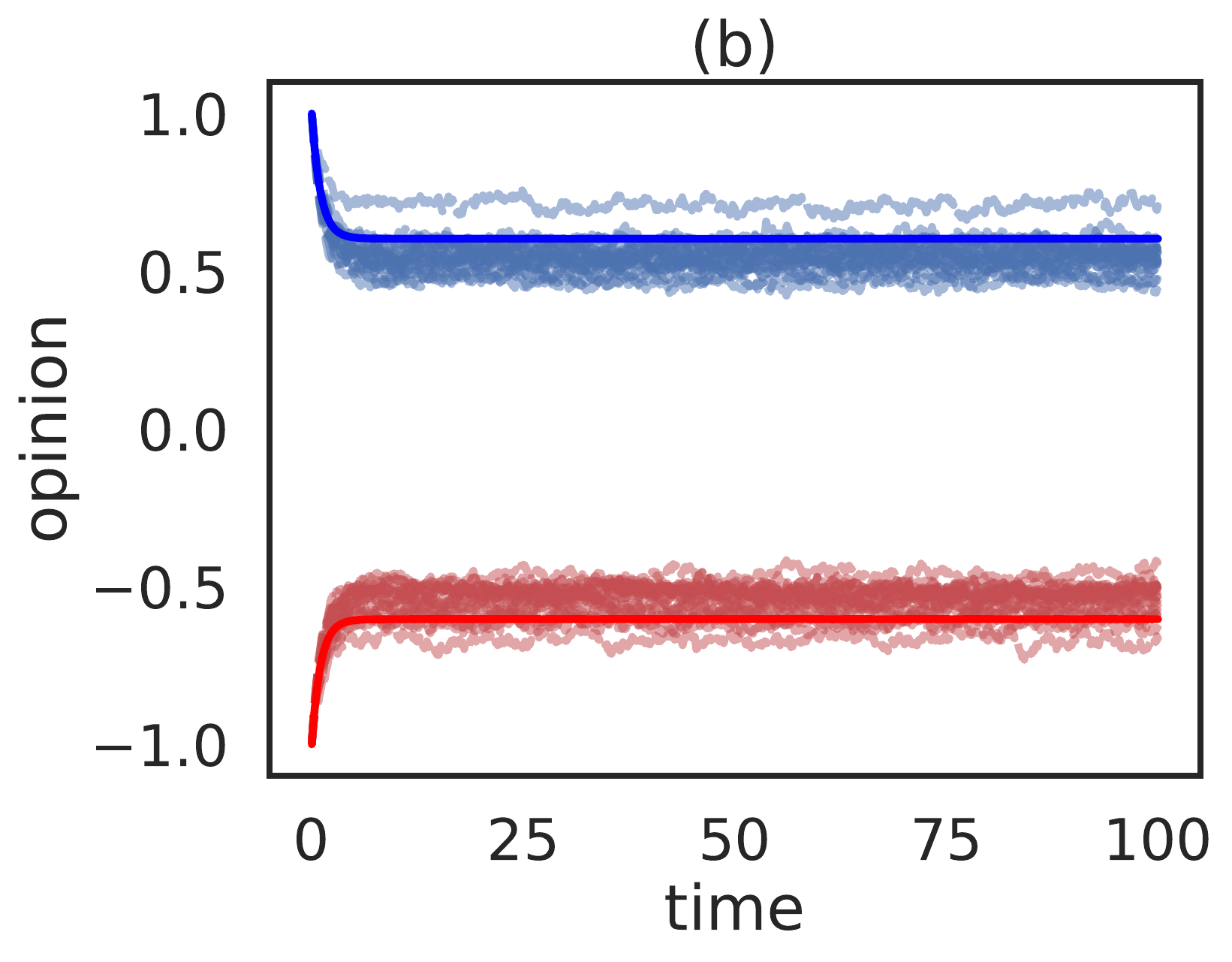}
\caption{Example trajectories of the groups' average opinions as they change in time. Dashed lines represent agent-based simulations and there are 20 independent realisations shown. Solid lines are the result of the mean field approximation (-MF). The top panel - (a) - shows the behaviour below the critical value with $\alpha = 0.84$ - both groups converge on a neutral opinion while the bottom panel - (b) - above it with $\alpha=4.0$ and groups remain in their respective opinions in opposition to each other.}
\label{fig:symmetric_trajectory}
\end{figure*}

\begin{figure*}[tb]
         \includegraphics[width=0.42\textwidth]{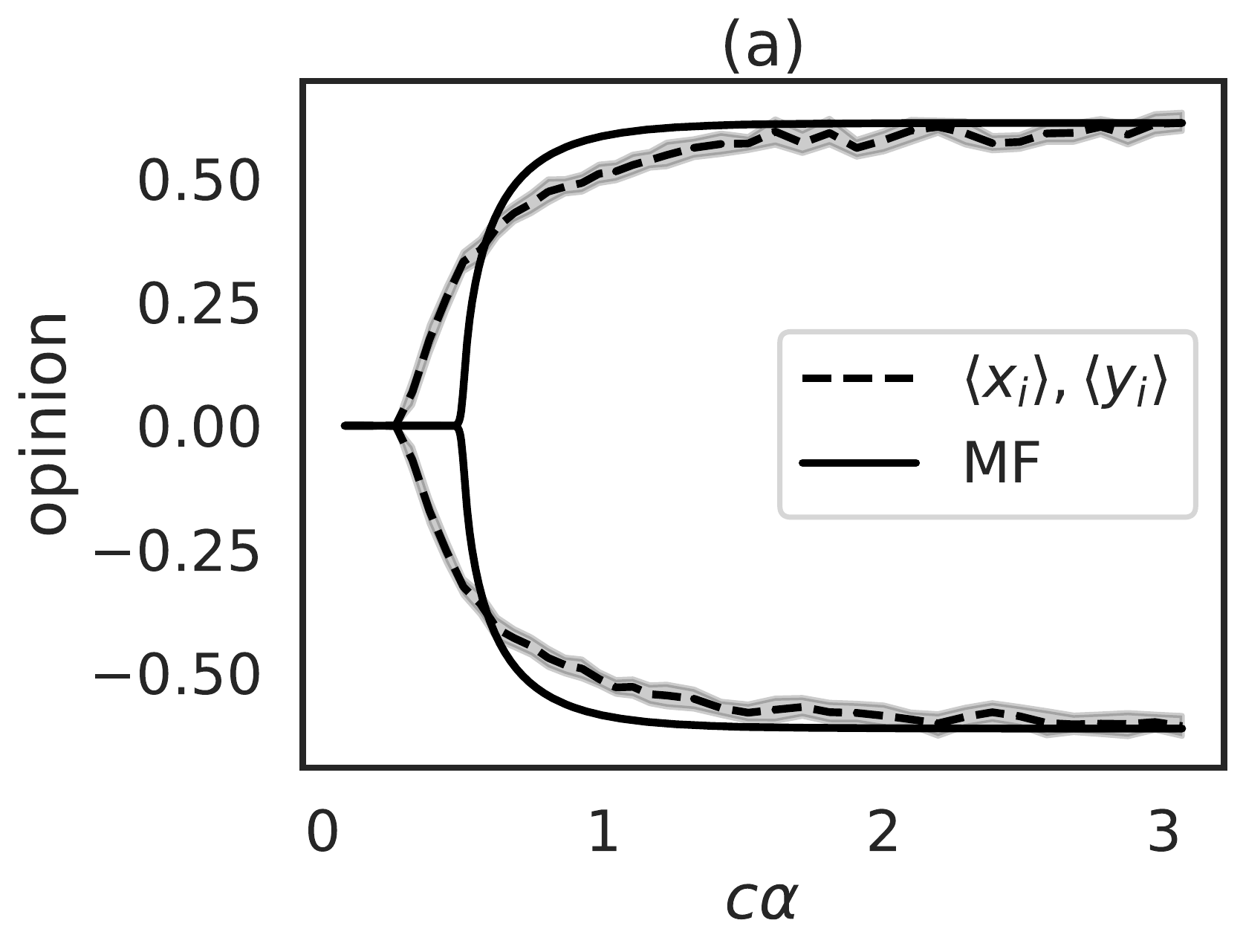}
         \hfill
         \includegraphics[width=0.42\textwidth]{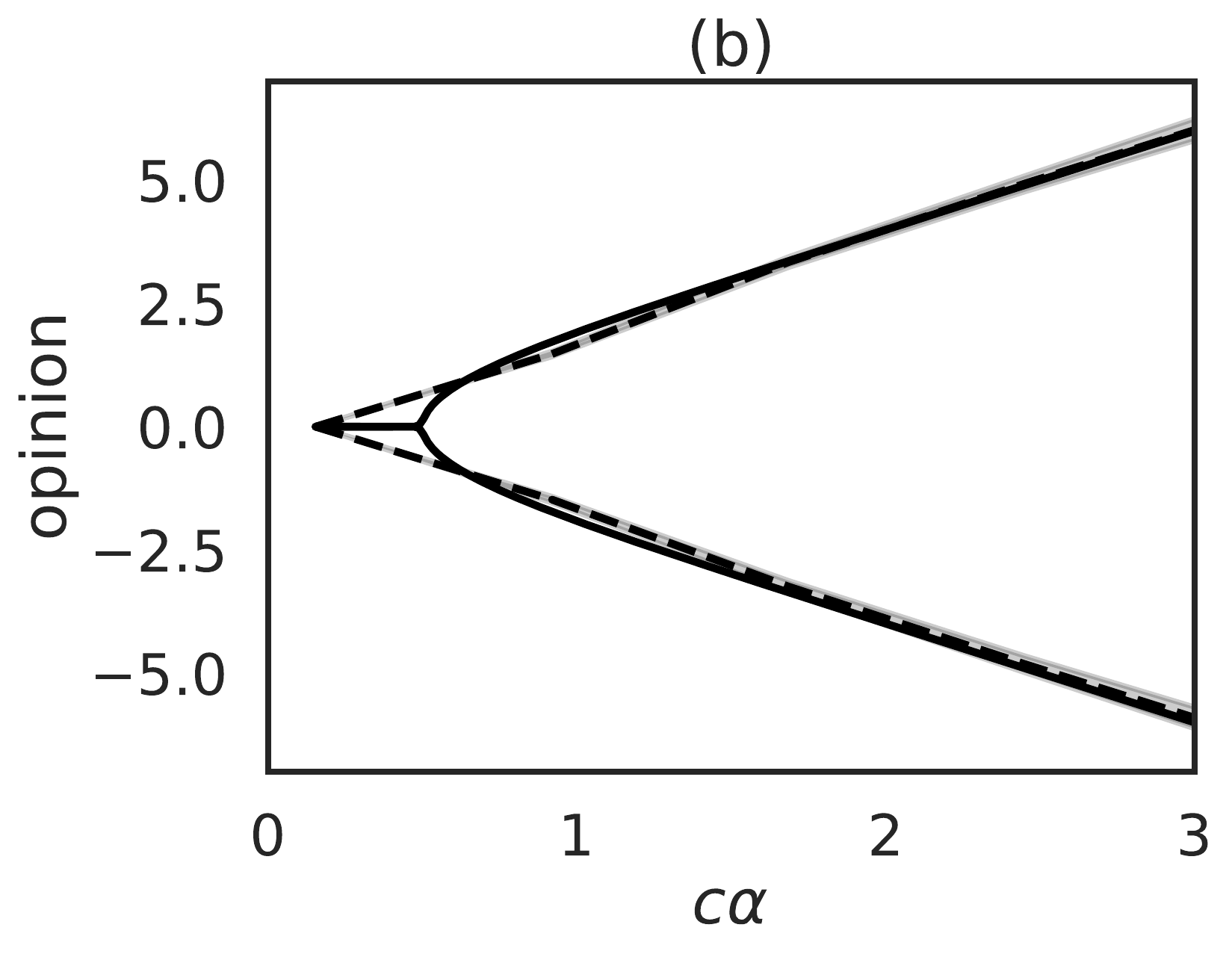}
\caption{Phase transition (a pitchfork bifurcation) from the symmetric consensus to the opposite  values  of opinions in different layers  $X$ and $Y$ under different parameter modulation in both agent-based simulations and mean field approximation. The transition takes place at the point  $K(1-\delta)c\alpha =1$.   The left panel - (a) - shows the transition as we increase $\alpha$ and keep  $c\approx0.306$  while the right panel - (b) - shows what occurs when we keep $\alpha=1$ and change $c$ by increasing the parameter $m$. The agent-based results are averaged over 20 independent realisations with a 95\% confidence interval present in the form of the error bands. Asymptotic behaviours observed at both panels for $c\alpha \gg 1$ are in a very good  agreement with Eq. (\ref{eq:xapprox}).}
\label{fig:sym_crit}
\end{figure*}

\begin{figure}[tb]
         \includegraphics[width=.38\textwidth]{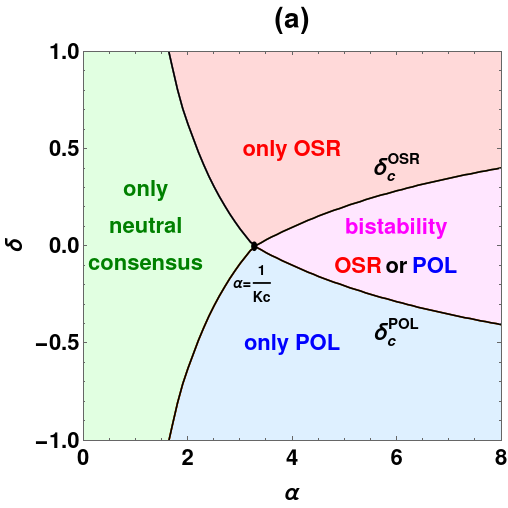}\\
        \includegraphics[width=0.45\textwidth]{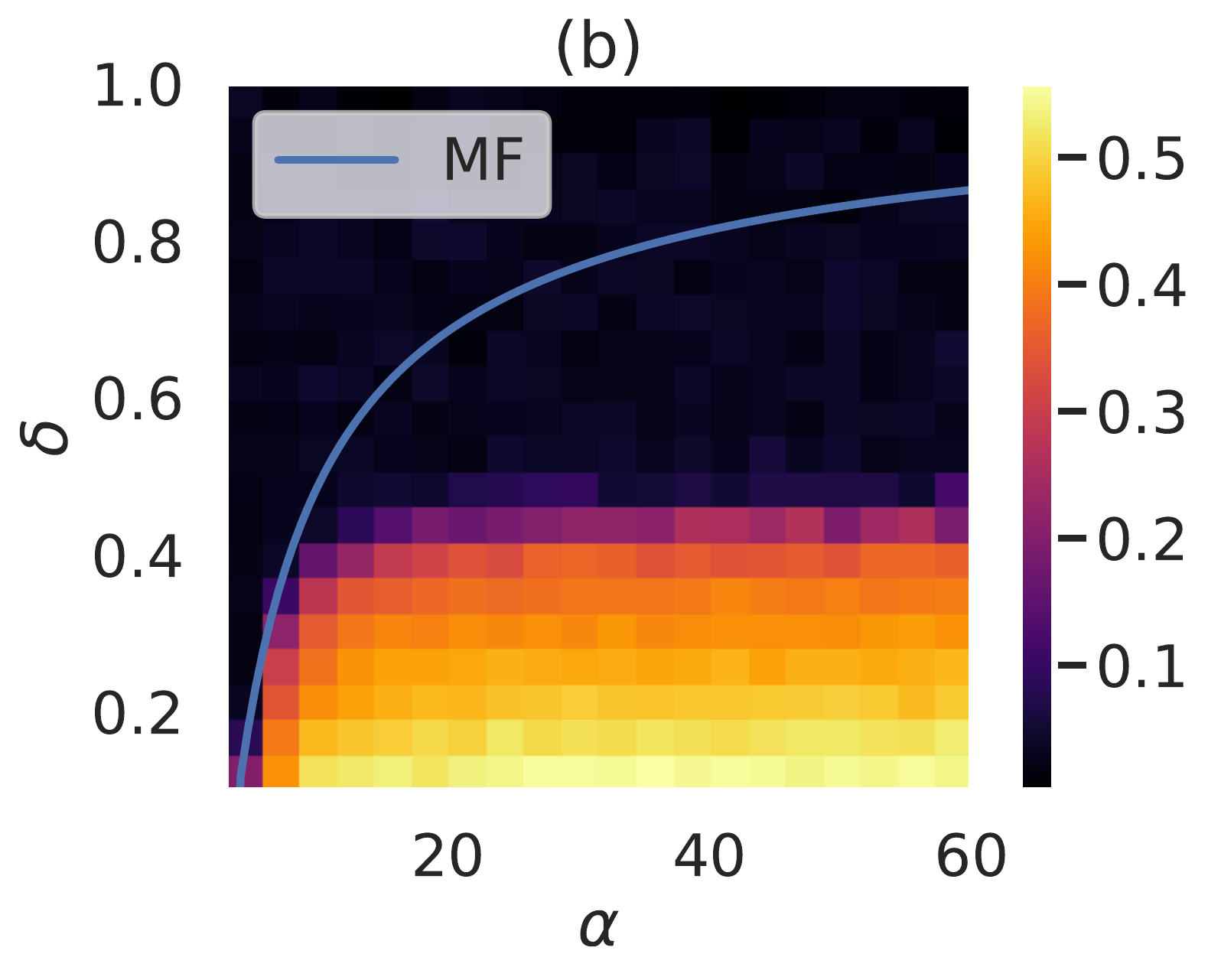}
\caption{(a) Phase diagram of the system given by Eq.~(\ref{eq:delta}) for $c=0.306$ and $K=1$. (b) Numerical simulations of the ABM model, where colour is $|\langle x \rangle - \langle y \rangle|$, with a visible transition from opposing opinions to a non-neutral consensus. In both panels solid lines come from MF solution given by Eq.~(\ref{eq:delta-crit}).}
\label{fig:scheme}
\end{figure}

Figure \ref{fig:weakscheme} illustrates this behavior via $(x, \delta)$ and $(x-y, \delta)$ planes plots with points (1)--(10) and, equivalently, (1') -- (10') referring to specific states of the system. The two layers start in opposition, i.e., in a polarised state ($\delta=0$, either 1 or 1', depending on the setting); then we enable a positive but weak $0 < \delta < 1$ coupling between them. As $\delta$ increases the groups final average opinions slowly and smoothly approach each other until the critical value of $\delta^{POL}_c$ (2 or 2') corresponding to a bifurcation point, where two groups merge into one with a radicalised opinion (3 or 3'). Further increasing of $\delta$ results in stronger radicalization (4 or 4'). On the other hand, if we follow the path of decreasing $\delta$ the average opinion value drops down (5, or 5') and we arrive once again at $\delta=0$ (6 or 6'). Although the value of $x$ at (6) is the same as in (1) it is a different state as confirmed by Fig. \ref{fig:weakscheme}b. We might then keep on decreasing $\delta$, switching to negative values (weak negative coupling) until we reach $\delta^{OSR}_c = -\delta^{POL}_c$ at (7 or 7') that once again corresponds to a bifurcation point, this time leading to separation of groups (8 or 8'), i.e. to a POL state. Further decrease of $\delta$ strengthens group polarisation (9 or 9') while by increasing it we go through (10 or 10') to close the loop reaching (1 or 1').  We also see a decent match of the mean field approach with agent-based model (ABM) simulations (Figure \ref{fig:weakscheme}c-d).   

We can interpret these results by posing a following question. Imagine that we can somehow influence the attitudes of the layers such that we soften the animosities towards more amicable, and maybe even eventually slightly cordial, side of things. Would that be enough to settle a conflict of some sort? Or do we need to completely flip peoples attitudes to make consensus possible. Our model suggests that it can be enough, indeed. This implies that while prejudice can cause society to split there is also room for hope because not as drastic changes to the attitudes as one would perhaps expect can cause the layers to converge on an opinion, albeit not a neutral one.

Figure \ref{fig:fig3D} presents solutions of Eq.~(\ref{eq:delta}) for different values of $\alpha$. It is essential to note here that in this system we face also other critical behaviour: in order to observe bi-stability for POL and OSR it is necessary that $\alpha > 1/(Kc)$ (see Fig.~\ref{fig:fig3D}a, c and e). Otherwise, if starting from a polarised state for $\delta < 0$, the average opinion in both groups decreases with increasing $\delta$ and when $\delta = -\frac{1}{Kc\alpha}+1$ a state of neutral consensus is achieved characterized by $x=0$ and $y=0$. The system stays in this state until $\delta = \frac{1}{Kc\alpha}-1$ where both groups simultaneously acquire the same non-zero opinion (OSR state). When $\alpha > 1/(Kc)$ we obtain also an auxiliary solution (marked by black dashed line in Fig.\ref{fig:fig3D}a-d), which is, however, always unstable and therefore plays no role in the dynamics.

Let us consider now in detail the case of the symmetrically and negatively coupled opposing layers with small values of $\alpha$ (i.e., the setting shown in Fig. \ref{fig:fig3D}e-h) and check it with the outcomes of ABM. 
With the use of a mean field theory we expect a phase transition from a neutral consensus - where both groups converge at zero - to a polarised state where the layers remain in their respective opinions in opposition to one another, as the control value $c\alpha$ is increased. We choose not to use one single control parameter as the behaviour of the system slightly changes depending on whether we modulate $c$ or $\alpha$.

We arrive at that prediction similarly as before, i.e., from the Jacobian matrix (\ref{eq:Jacob_sym}) of the system (\ref{eq:delta}) we can acquire the eigenvalues  $\lambda_{+,-}$
\begin{equation}
  \lambda_{+,-}=c\alpha K (1 \pm \delta)-1.
\end{equation}

We can then find a steady state solution in the polarised phase ($x_{t\rightarrow\infty} = -y_{t\rightarrow\infty}$) by solving numerically  the following relation:
\begin{equation}
    x_{t\rightarrow\infty} = (1 - \delta) Kc \tanh{\alpha x_{t\rightarrow\infty}},
\end{equation}
which can be written in a normalised form   $u=K(1-\delta)\alpha c\tanh(u)$   when  $u=\alpha  x_{t\rightarrow\infty}$. Since the solution  $u$ of the last equation  increases from $0$ to   $K(1-\delta)\alpha c$  when the product $K(1-\delta)\alpha c$ increases from $1$ to $\infty$ thus for $K(1-\delta)\alpha c \gg 1$ there is 
\begin{equation} 
x_{t\rightarrow\infty}\approx K(1-\delta)c,
\label{eq:xapprox}
\end{equation} 
which explains the difference in the behaviour we mentioned ($c$ vs. $\alpha$ modulation) and observe in Fig.~\ref{fig:sym_crit}.

In Fig.~\ref{fig:symmetric_trajectory} we present examples of trajectories of the system where we arbitrarily chose groups to start with all its agents with opinion +1 (X) and with -1 (Y), however, the results do not depend on this choice.
There one can see the two aforementioned phases - consensus and polarisation. Plots show the mean opinion of each layer as a function of time. The agent-based simulations are not deterministic and therefore we show 20 independent realisations and compare against the mean field prediction. It is apparent that below the critical value of $c\alpha$ the whole system converges at zero - both layers reach a neutral consensus (Fig.~\ref{fig:symmetric_trajectory}a). As the control parameter is increased the situation changes and a polarisation phase occurs (Fig.~\ref{fig:symmetric_trajectory}b). The two layers now stand in opposition to one another and no consensus is possible.

Using the mean field theory we estimate the critical value of $c\alpha$ and present the test of our predictions in Fig.~\ref{fig:sym_crit}. As mentioned before it depends whether we modulated $\alpha$ or $c$ and we show that in Fig.~\ref{fig:sym_crit}a and Fig.~\ref{fig:sym_crit}b, respectively. When $c=const.$ the system reaches a plateau, however, when $c$ is increased the final opinion value of the system also increases indefinitely. In both scenarios we see a phase transition (a supercritical pitchfork bifurcation \cite{strogatz2018nonlinear}) at a certain critical value and a reasonably decent fit from the mean field approximation. 

We find this setting to be representative of a typical echo chamber situation in context of two rivalling groups such as political parties. If the animosity from one to the other or mutually is strong enough then no consensus is possible - while the groups may not be as radical as in initially they will always persist in their view opposite to the other. This essentially shows that prejudice has the potential to lock society into a predetermined antagonistic state.

The outcomes of this analysis can be summarised in a concise way with a $\delta-\alpha$ phase diagram shown in Fig. \ref{fig:scheme}a where the predictions of different states of the systems (i.e, neutral consensus, polarization, radicalization and bi-stability) are presented. We also see a decent match of the ABM results, however, only for relatively small values of $\alpha$ and $\delta$ - see Fig.~\ref{fig:scheme}b where we show a heatmap of the $\delta-\alpha$ phase space where colour denotes the distance between averages.  

\subsection{Non-symmetric coupling}
\label{sec:non-symm}
Let us now make a bridge between the systems (\ref{eq:delta}) and (\ref{eq:system}) by formulating predictions in the mean field approach for the case when the coupling between layers is not symmetrical. At first we shall look at a scenario when the coupling is of the same sign but of (possibly but not necessarily) different magnitude, i.e., we consider the system as described by (\ref{eq:system}) with the omission of the external bias. The procedure of the analysis for this systems is, of course, analogous to what we have already done before:

\begin{figure*}[htb]
         \includegraphics[width=0.42\textwidth, height=0.265\textheight]{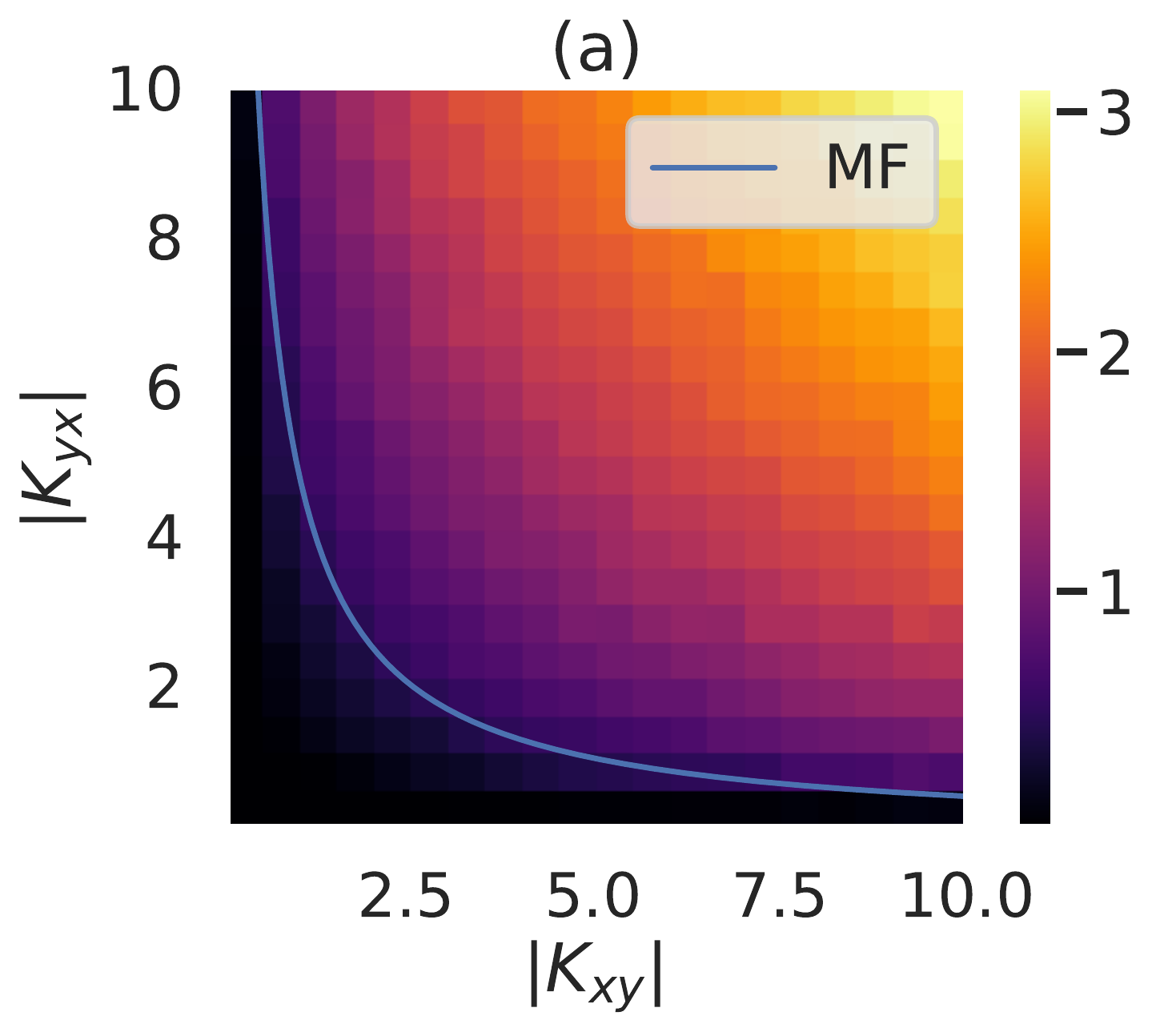}
         \hfill
         \includegraphics[width=0.42\textwidth]{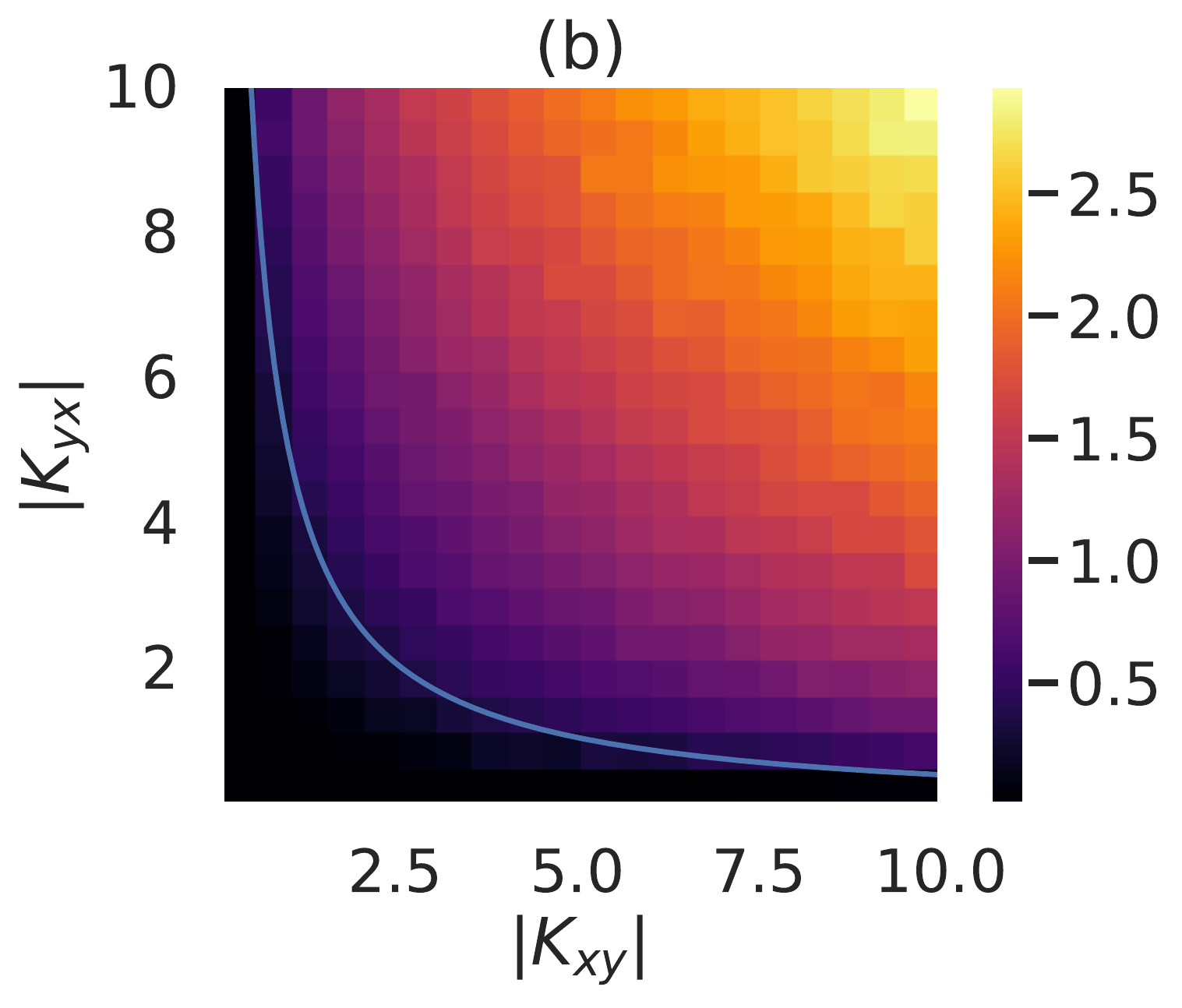}
\caption{The coupling parameters phase space ($|K_{xy}|-|K_{yx}|$, when $K_{xy}, K_{yx}<0$ in a form of a heat map, where the colour represents the average  $(1/2)(|\langle x_i \rangle|+|\langle y_i \rangle|)$, with a visible transition from neutral consensus to polarisation. MF is Eq.~(\ref{eq:curve}). Left panel - (a) - shows results for initial conditions corresponding to the opposite radicalisation in each layer. In the right panel - (b) - the initial conditions for all agents were drawn randomly from a uniform distribution $(-1, 1)$ showing that this result does not depend on initial conditions.}
\label{fig:sym_crit_heat}
\end{figure*}

The Jacobian matrix of (\ref{eq:system}) is
\begin{equation}
J|_{\langle x \rangle=\langle y \rangle=0} = 
\begin{bmatrix}
 -1 + c\alpha K & c\alpha K_{xy} \\
 c\alpha K_{yx} & -1 + c\alpha K
\end{bmatrix},
\label{eq:Jacob}
\end{equation} 
from which we get both eigenvalues as:
\begin{equation}
    \lambda_{1,2} = c \alpha K \mp c \alpha \sqrt{K_{xy}K_{yx}} - 1.
\end{equation}
When $K_{xy}=K_{yx}$ then eigenvalues $\lambda_{1,2}$ reduce to $\lambda_{+,-}$
\begin{equation}
   \lambda_{+,-}=c\alpha(K \pm K_{xy})-1,
\end{equation}
calculated directly from the agent-based model (\ref{eq:2groups}) in the limit $N\to \infty$ and in such a case corresponding eigenvectors of Jacobian (\ref{eq:Jacob}) are $\boldmath{e}_+=[1,1]^T$ and  $\boldmath{e}_-=[1,-1]^T$. 

In general  the product $K_{xy}K_{yx}$ can be positive or negative; if it is positive then either $K_{xy}>0 \land K_{yx}>0$ and the system falls into what was described by Baumann et al. (unless we consider the weak coupling $\delta < 1$ introduced in Sec.~\ref{sec:sym})
or $K_{xy}<0 \land K_{yx}<0$ and new behaviour in the system emerges accompanied by a phase transition occurring when $\lambda_{max} = \lambda_{-}$ changes sign. Since the eigenvector $e_-$ is asymmetrical thus the case  $\lambda_{max}>0$  means here  that  the consensus phase $x=y=0$ looses its stability and systems is  polarised, i.e. opinions in groups $X$ and $Y$ split into opposite directions.  
From $\lambda_{max}$ changing its sign we get a relationship between the $K_{xy}$ and $K_{yx}$:
\begin{equation}
    K_{yx} = \bigg(\frac{1-c\alpha K}{c\alpha}\bigg)^2  \frac{1}{K_{xy}}.
    \label{eq:curve}
\end{equation}
If $K_{yx} = K_{xy}<0$ and the system is in the polarised phase then its steady state is $x_{t\rightarrow\infty} = -y_{t\rightarrow\infty}$ which can be found by solving numerically for $x_{t\rightarrow\infty}$ the following relation:
\begin{equation}
    x_{t\rightarrow\infty} = (K-K_{xy})c~\tanh(\alpha x_{t\rightarrow\infty}).
    \label{xassympt}
\end{equation}
\begin{figure*}[htb]
         \includegraphics[width=0.32\textwidth]{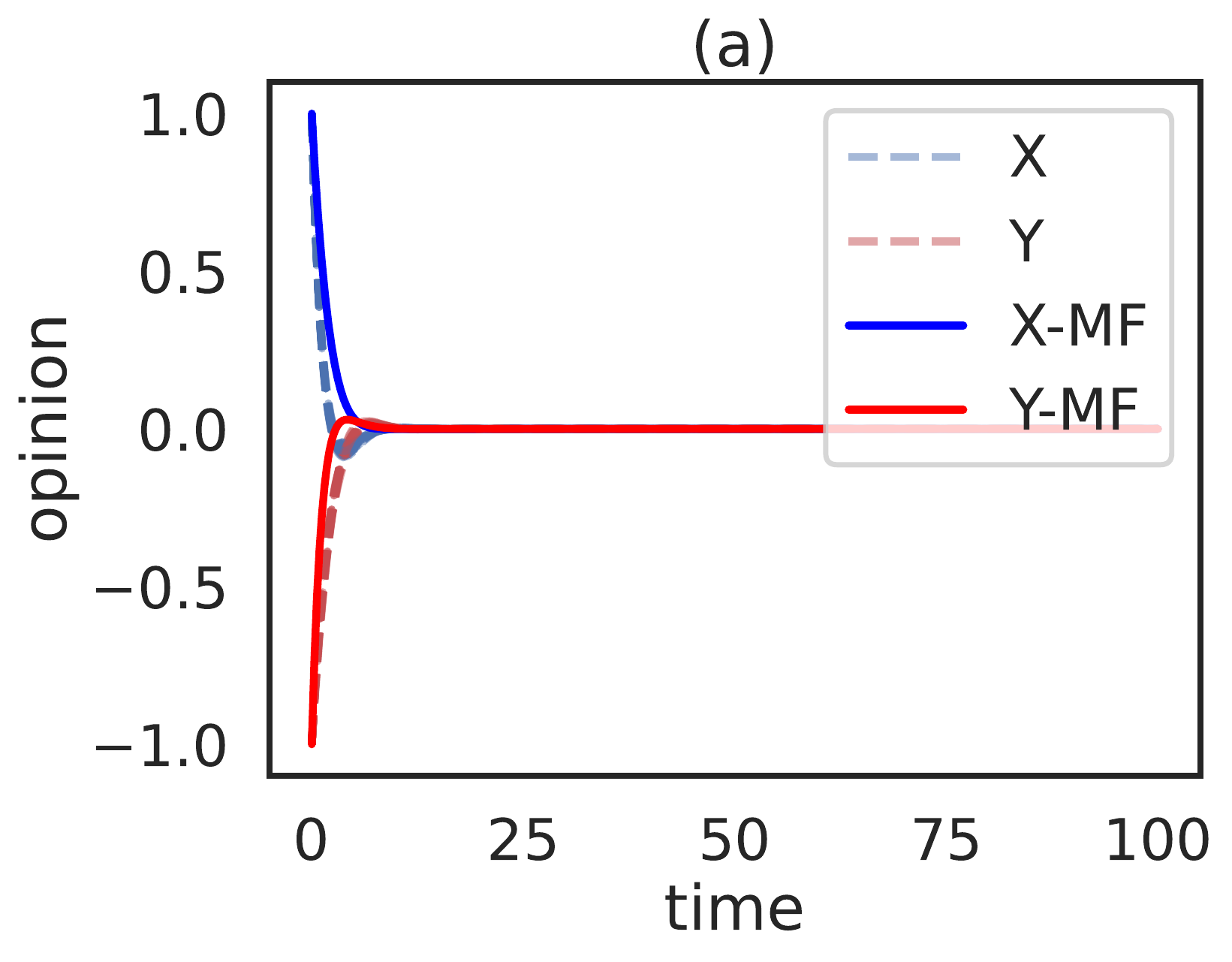}
         \hfill
         \includegraphics[width=0.32\textwidth]{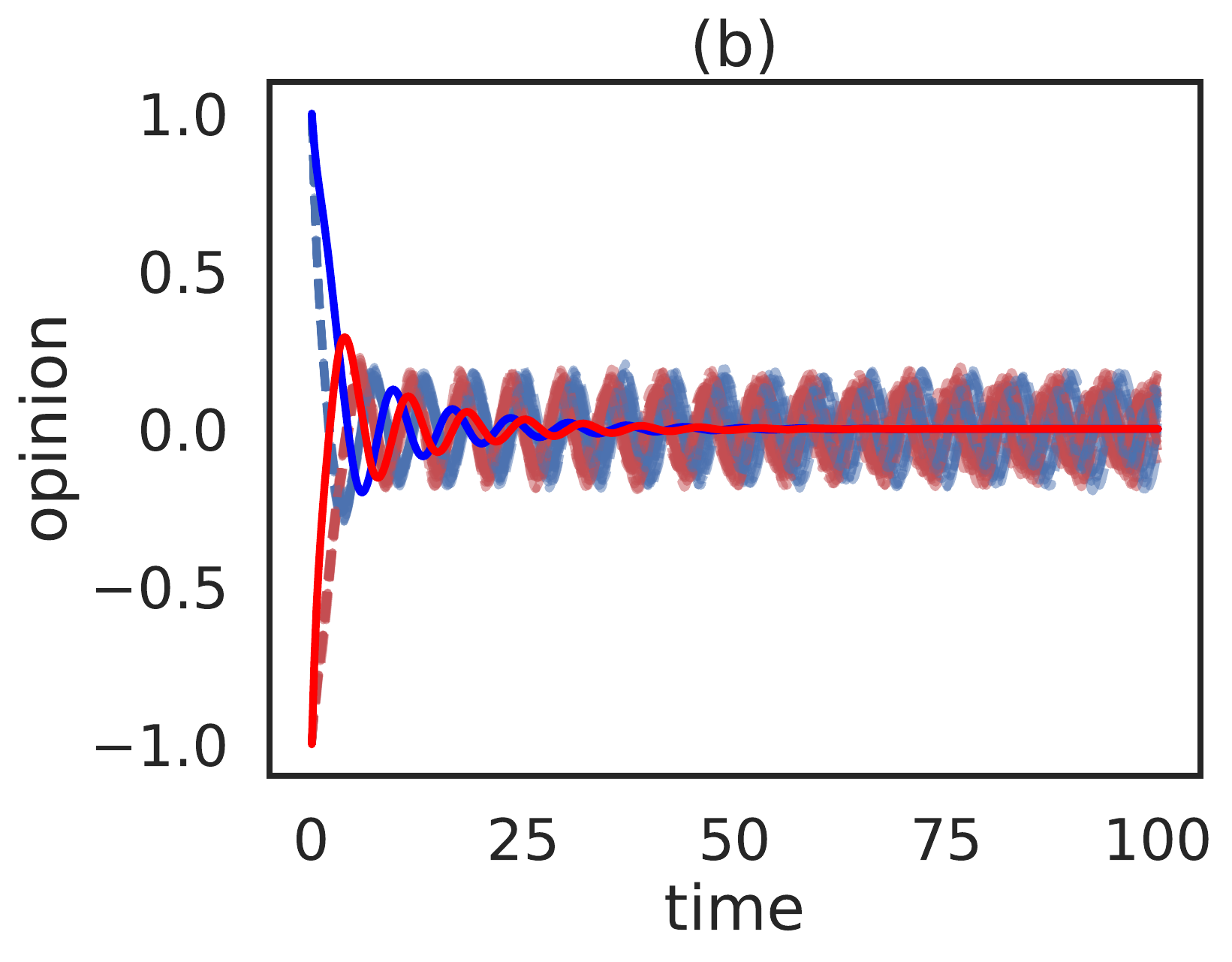}
         \hfill
         \includegraphics[width=0.32\textwidth]{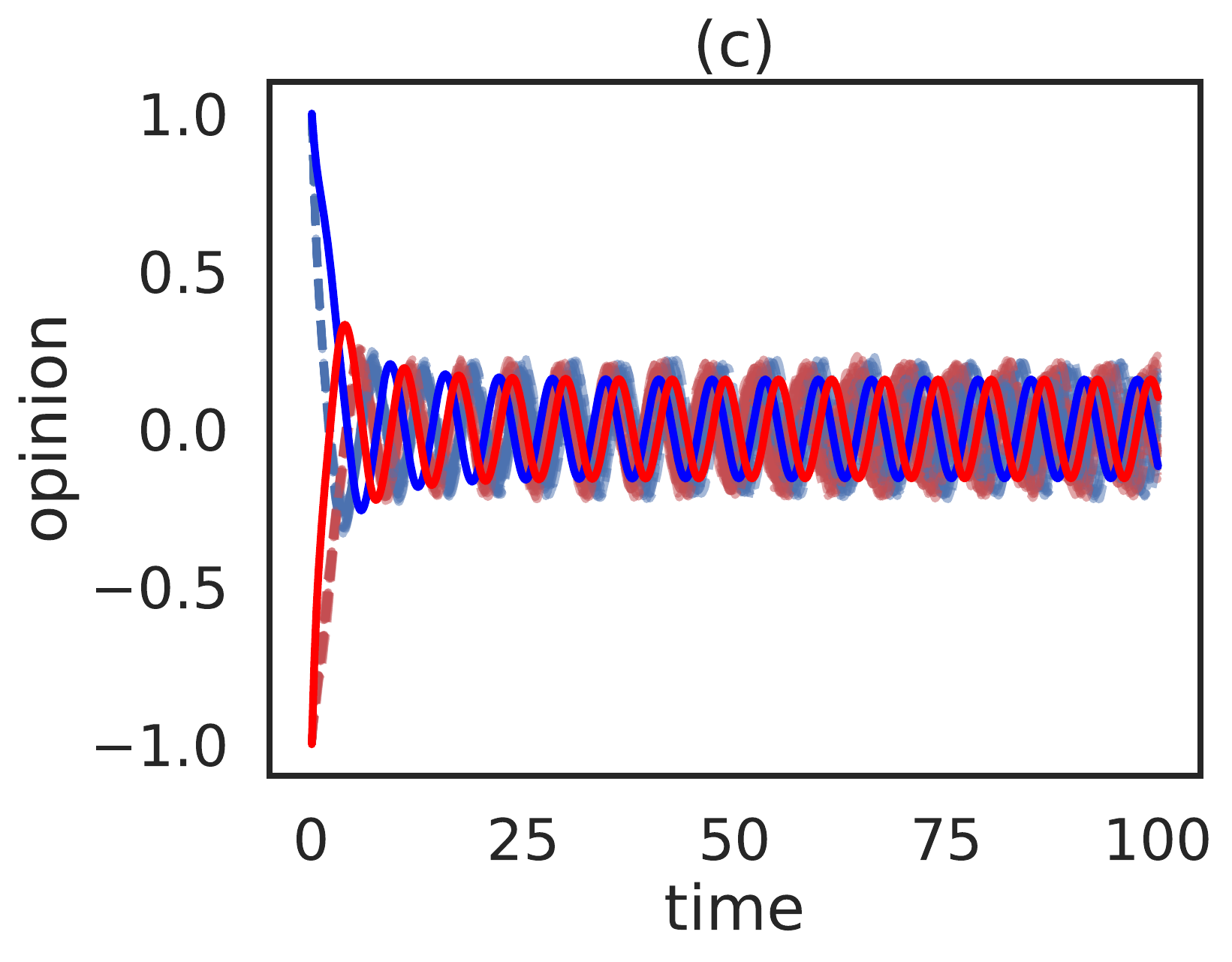}
\caption{Example trajectories in the asymmetric coupling parameters scenario for $\alpha=1,~3,~3.5$ respectively left to right. 20 independent agent-based simulation results are shown as dashed lines with solid lines representing the mean field (MF) approximation. Two distinct behaviours are visible - sustained and dampened oscillations.}
\label{fig:asymmetric_trajectories}
\end{figure*}

Equation~(\ref{xassympt}) can be again written in a normalised form as earlier  $u=(K-K_{xy})\alpha c\tanh(u)$   when  $u=\alpha  x_{t\rightarrow\infty}$. Since the solution  $u$ of the last equation  increases from $0$ to   $(K-K_{xy})\alpha c$  when the product $(K-K_{xy})\alpha c$ increases from $1$ to $\infty$ thus for $(K-K_{xy})\alpha c \gg 1$ there is 
\begin{equation} 
x_{t\rightarrow\infty}\approx (K-K_{xy})c,
\label{eq:xapprox2}
\end{equation} 
which also explains the difference in the behaviour we observe in Fig.~\ref{fig:sym_crit} albeit in a more general context.

We also present a heatmap (Fig.~\ref{fig:sym_crit_heat}) of the coupling parameters phase space with $Kc\alpha \approx 0.306$. The colour there shows the absolute value of the mean opinion of the system. Again we see a transition from consensus to polarisation with a good match from the mean field approach and specifically the Eq.~(\ref{eq:curve}).

\begin{figure*}[htb]
         \includegraphics[width=0.43\textwidth]{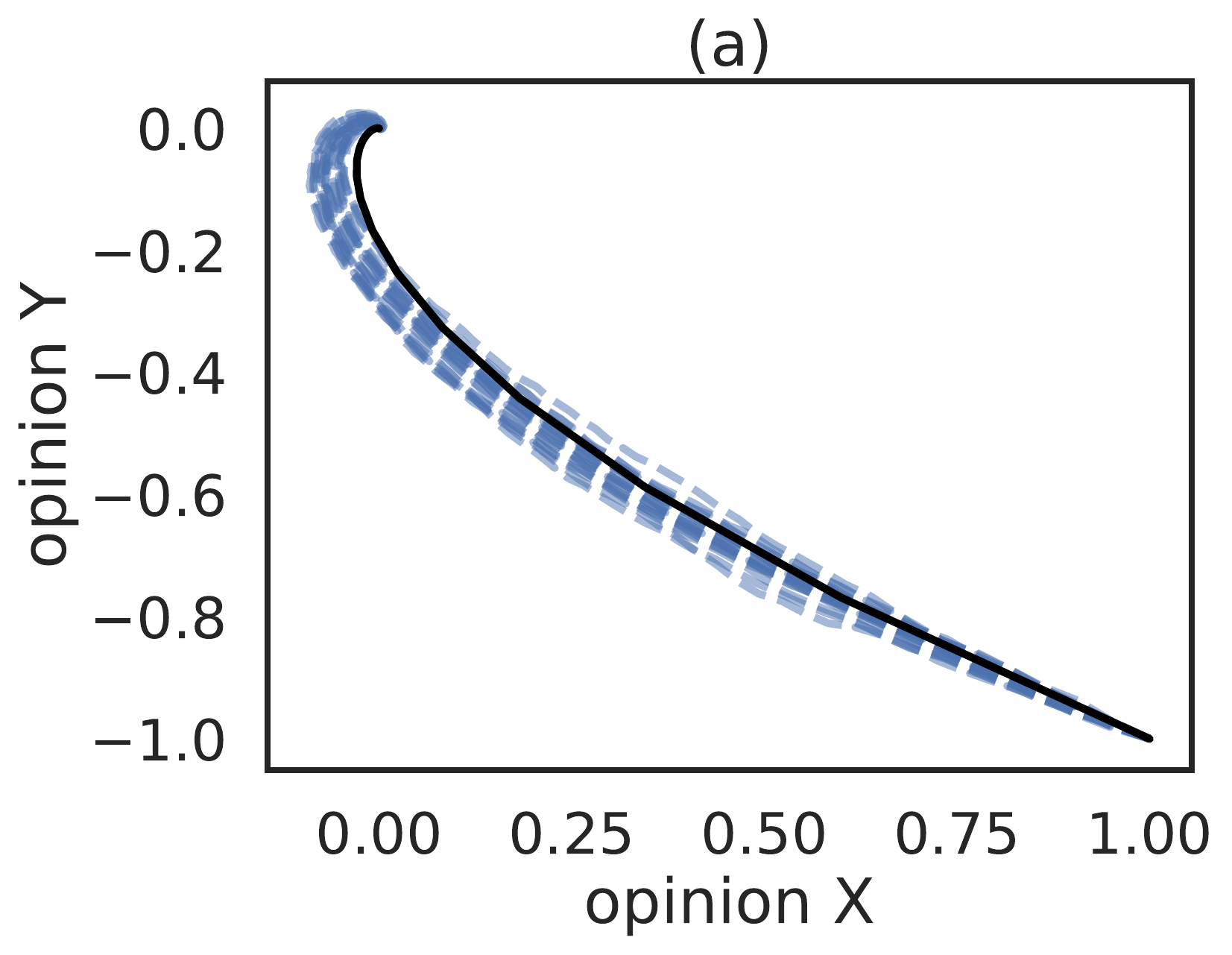}
         \hfill
         \includegraphics[width=0.43\textwidth]{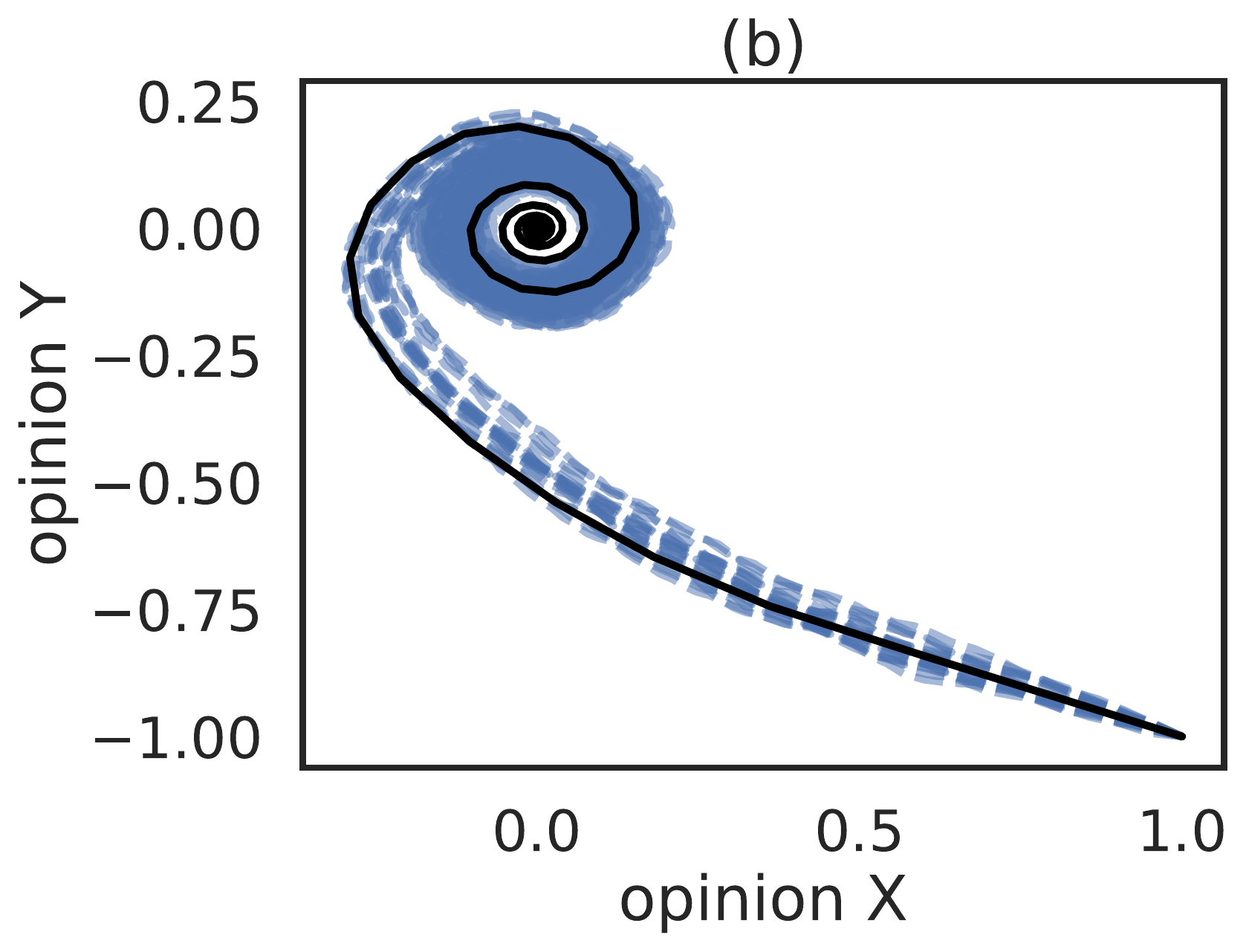}
         
         \includegraphics[width=0.43\textwidth]{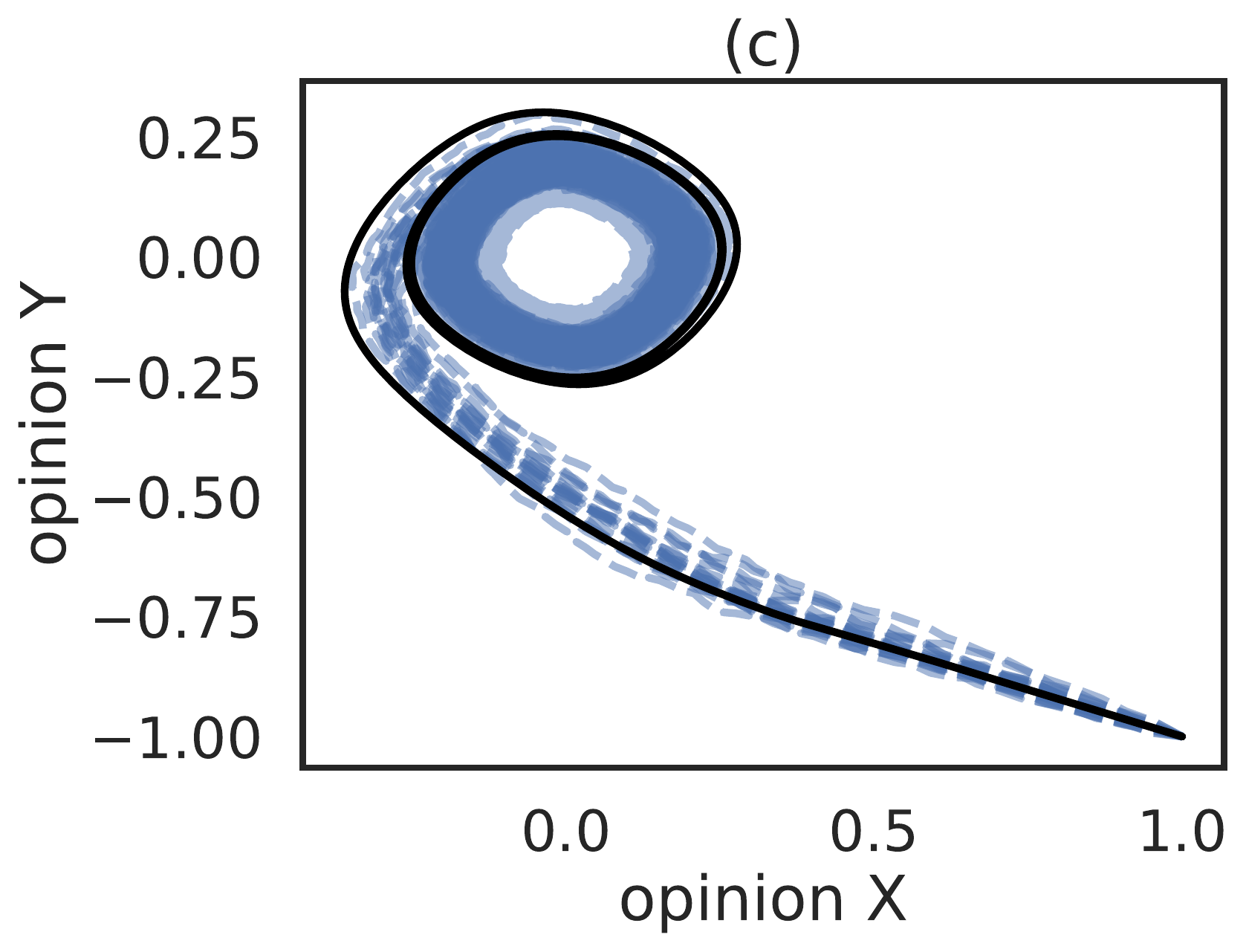}
         \hfill
         \includegraphics[width=0.43\textwidth]{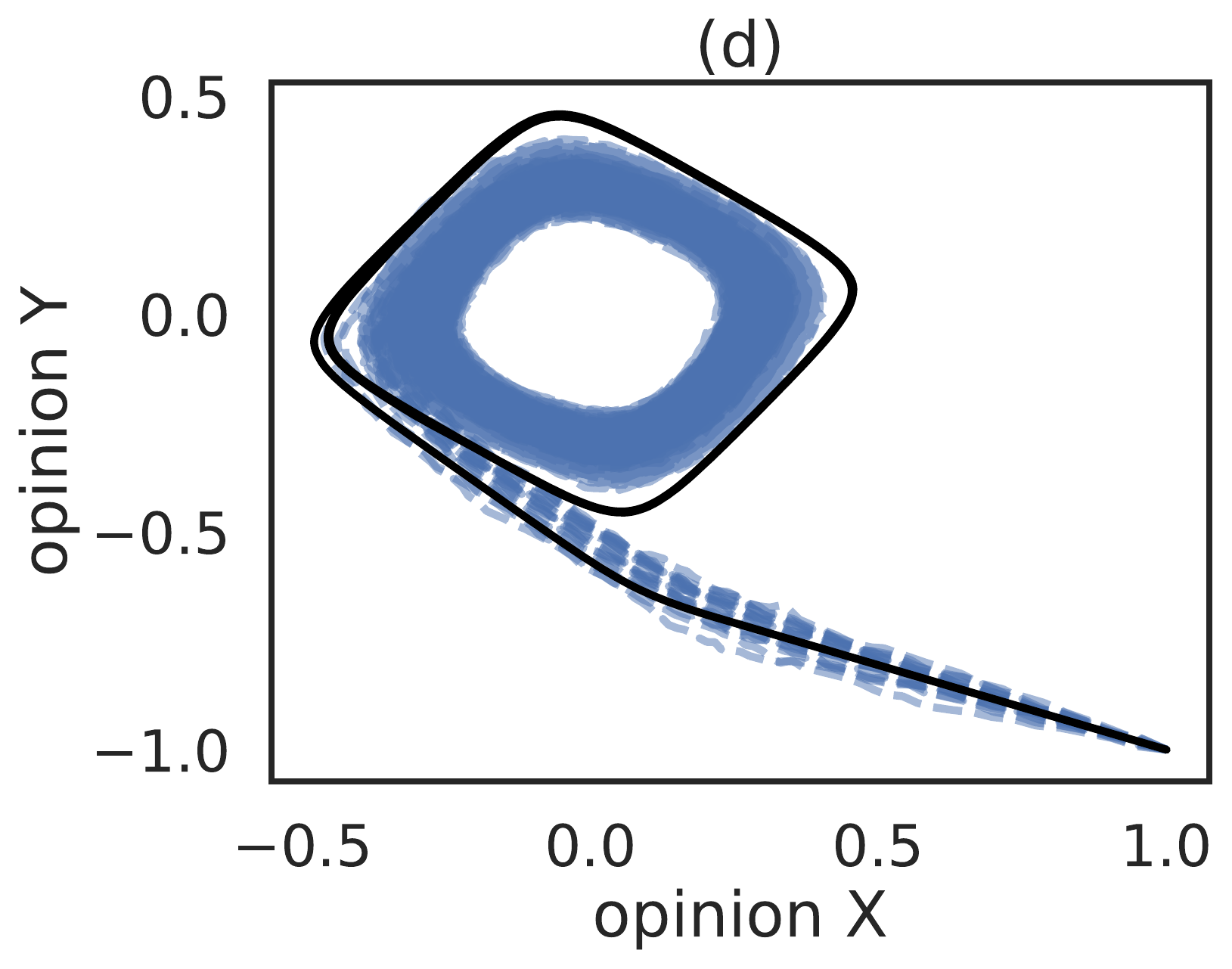}
\caption{Y(X) trajectories in the asymmetric coupling scenario for $\alpha=1,~3,~4,~10$ at (a), (b), (c), and (d) respectively, with dashed lines representing 20 independent realisations of the agent-based simulation and solid lines showing the mean field solution. We observe in detail that the system has two possible attractors - a point and an orbit. For $Kc\alpha > 1$ the point (0,0) becomes unstable and trajectories starting from it would also end in an orbit.}
\label{fig:asymmetric_xy_trajectories}
\end{figure*}

Another interesting case is that of an a- or perhaps even anti- symmetric coupling where one group ``likes'' the other but the feeling is not mutual, i.e., the signs of the coupling parameters are opposite. According to the mean field theory we ought to see two possible behaviours of the system - dampened or sustained oscillations depending on the values of the control parameter. As before it does depend whether we change $c$ or $\alpha$. In Figs.~\ref{fig:asymmetric_trajectories} and ~\ref{fig:asymmetric_xy_trajectories} we show time and phase trajectories respectively. In both cases it is apparent that the two aforementioned behaviours are present. Namely the system has two possible attractors - a point or an orbit. While there is a slight shift as to when the transition occurs when comparing agent-based simulations and the mean field approximation we find the analytical approach to be qualitatively successful.

\begin{figure*}[htb]
     \includegraphics[width=0.43\textwidth]{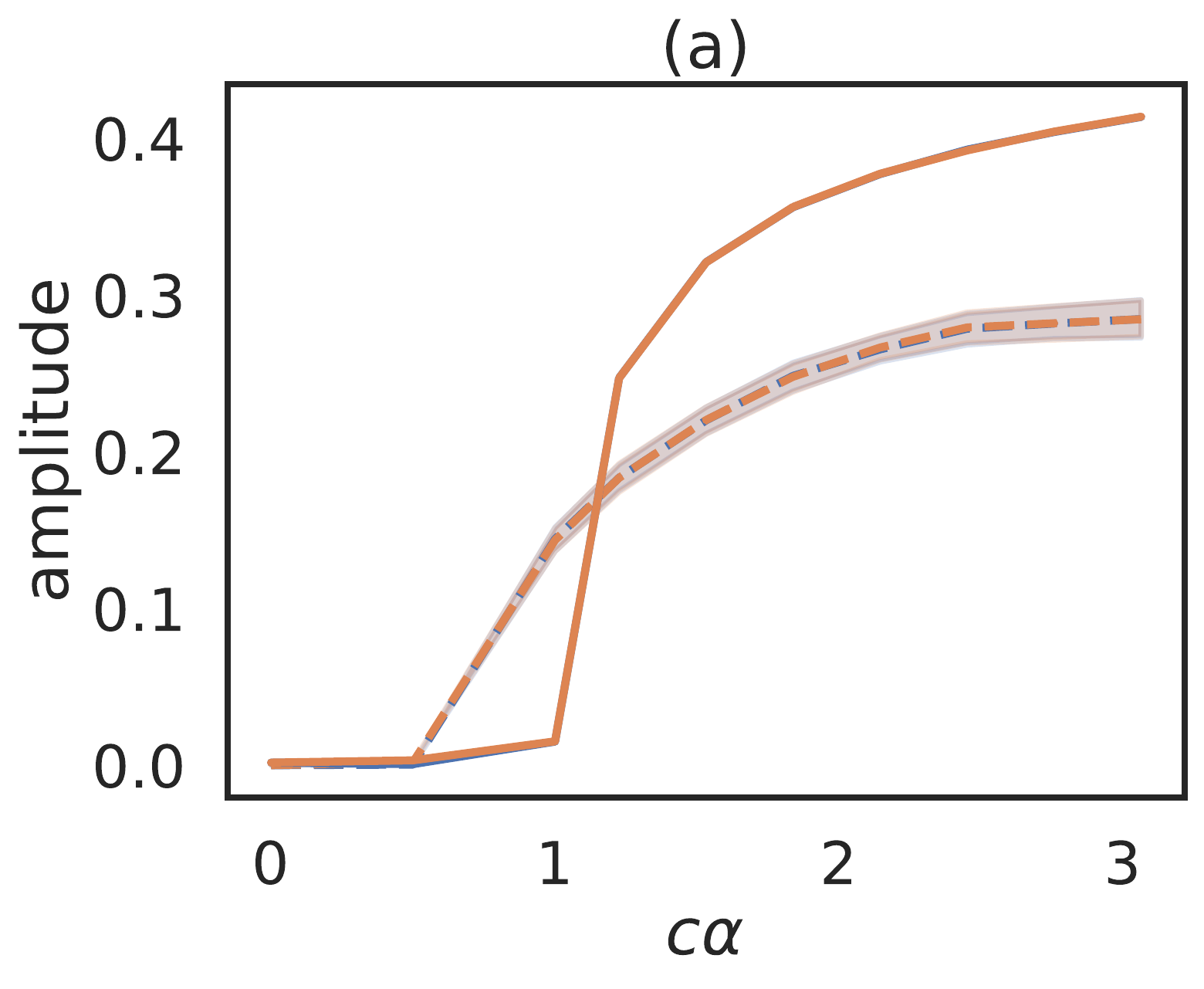}
     \hfill
     \includegraphics[width=0.43\textwidth]{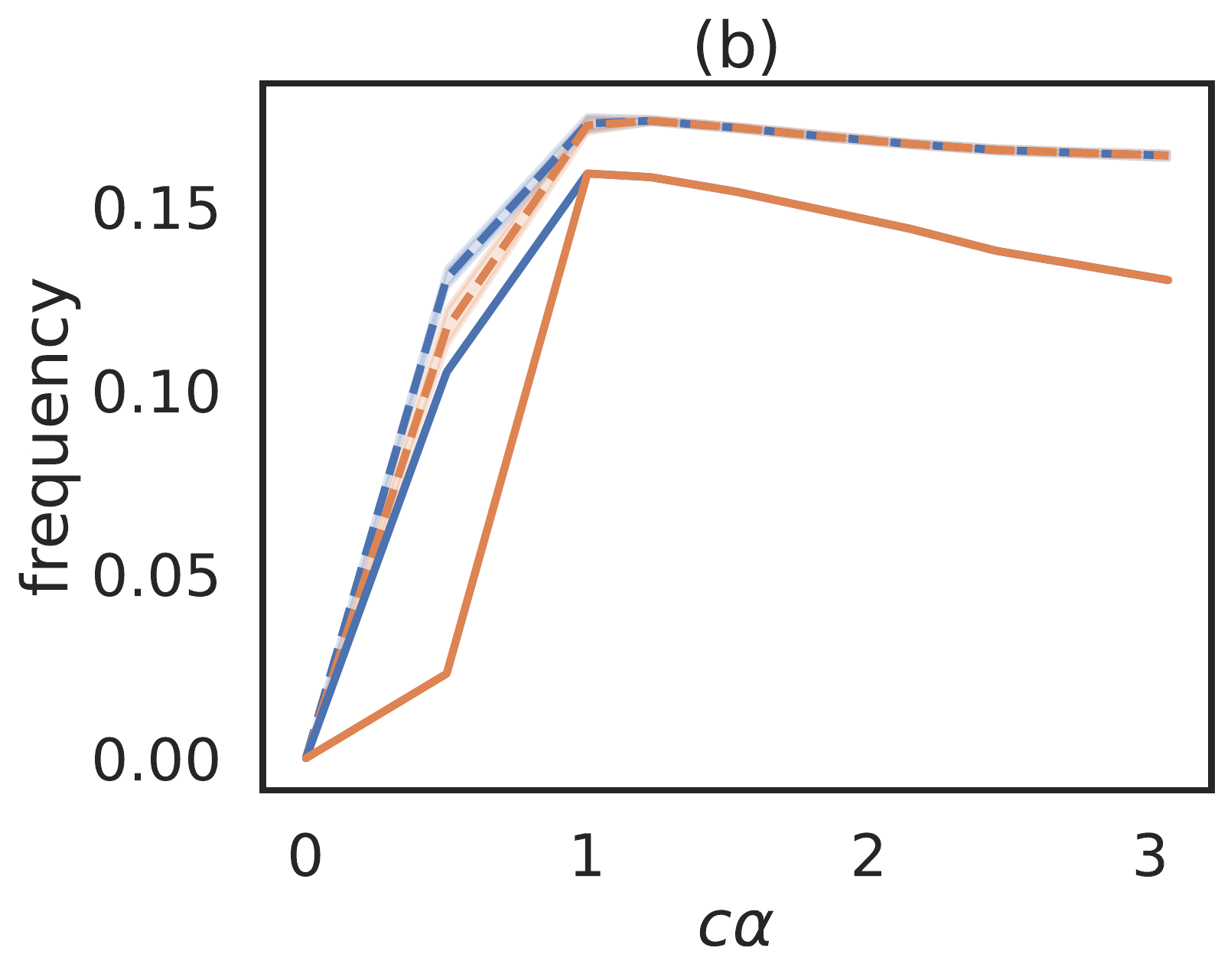}
     
     \includegraphics[width=0.43\textwidth]{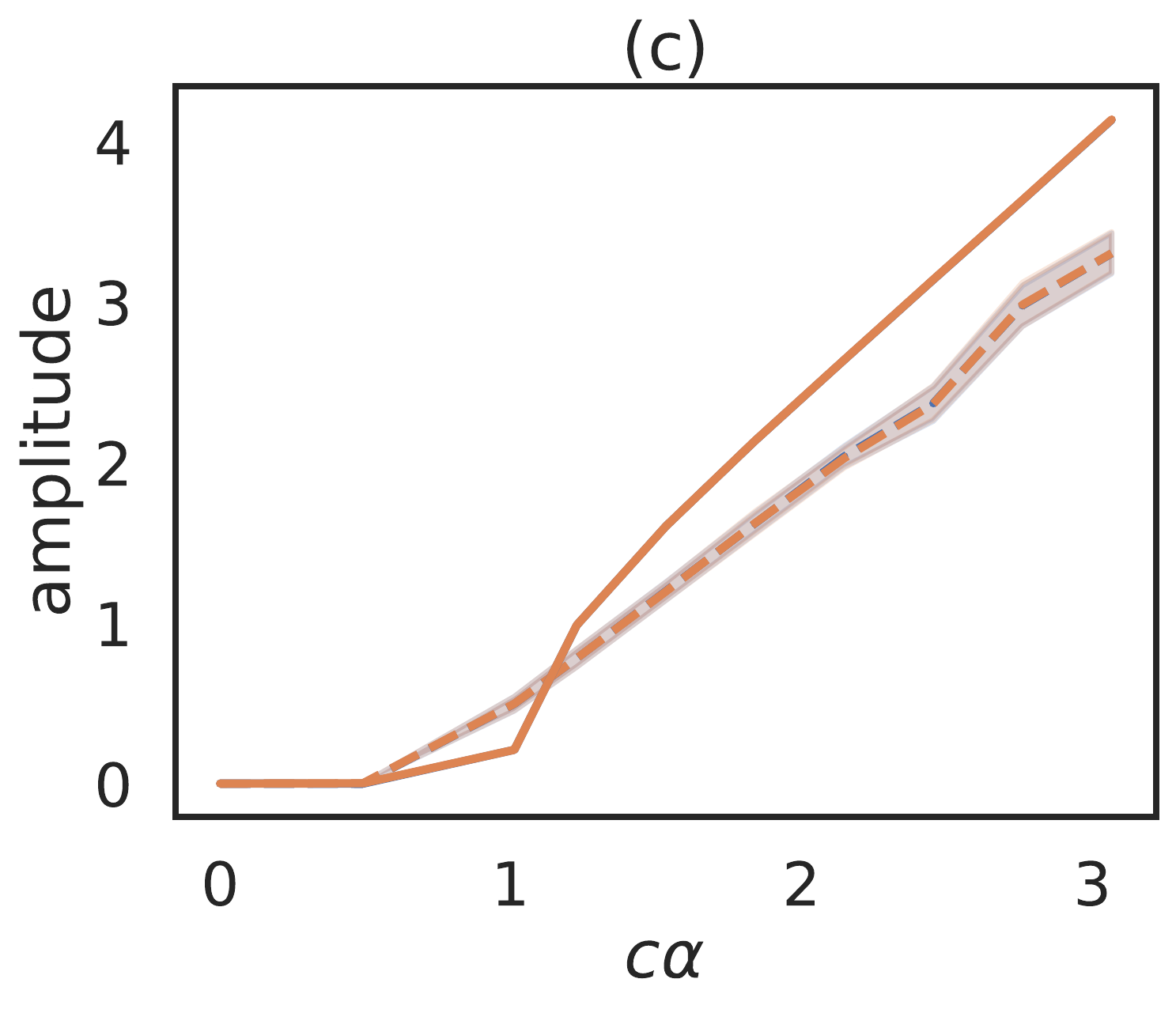}
     \hfill
     \includegraphics[width=0.43\textwidth]{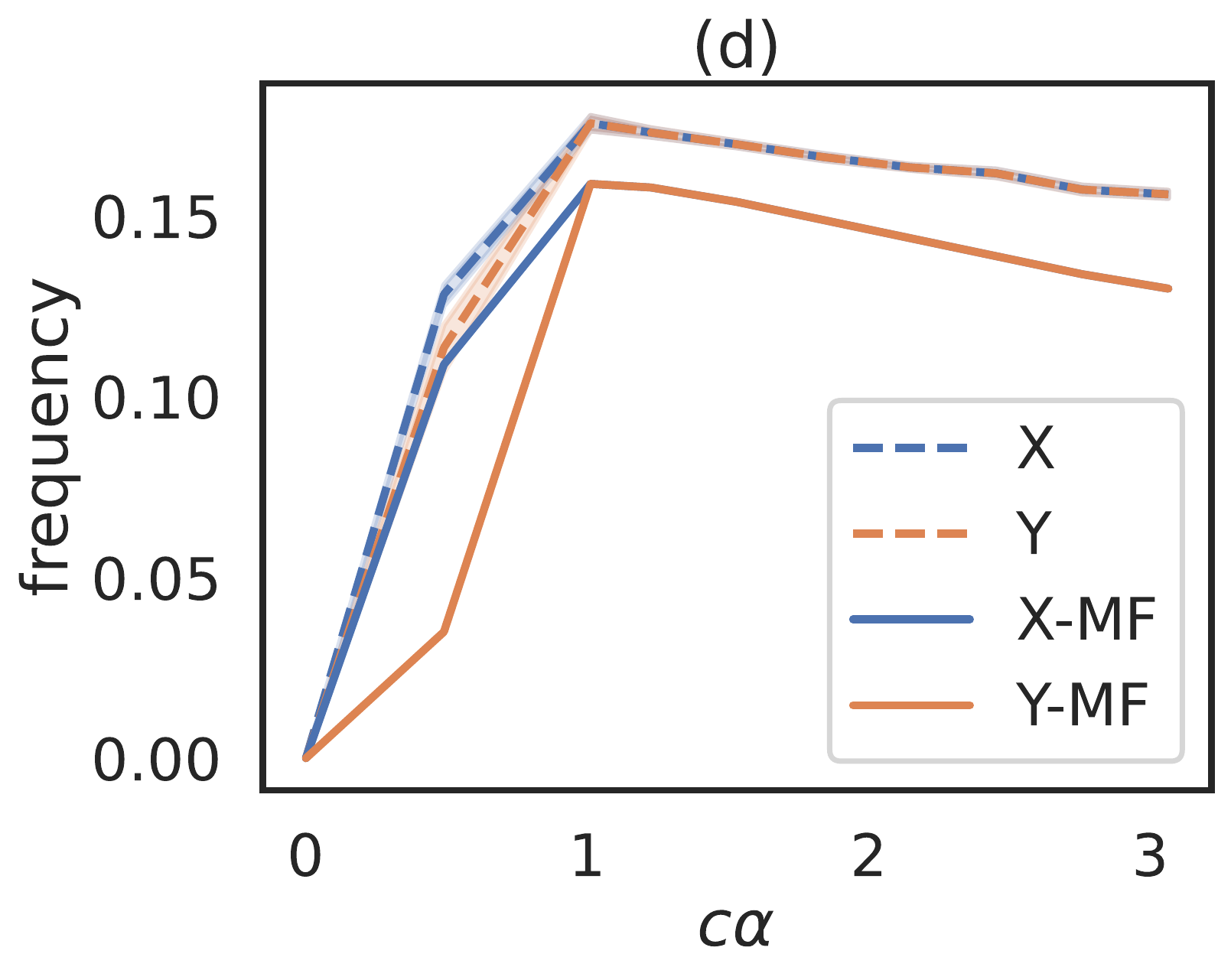}
\caption{The oscillations frequencies and amplitudes dependence on the parameter modulation in the asymmetric coupling case. Top row is for $\alpha$ modulation and the bottom one for $m$. Dashed lines represent an average over 20 independent realisations of agent-based simulations with a 95\% confidence interval present as the error bands. Solid lines show the mean field solution. Differences in asymptotic behaviours of amplitudes in (a) and (c) panels are similar to differences observed for values of the steady state solutions $x_{t\to \infty}$ at the left and right panel of Fig. \ref{fig:sym_crit}.}
\label{fig:asymmetric_freq}
\end{figure*}

This effect is due to the product $K_{xy}K_{yx}$ being negative and then the eigenvalues are complex, and the system exhibits a supercritical Hopf bifurcation \cite{strogatz2018nonlinear}. When $Kc\alpha <1$ then the attractor of dynamical system (\ref{eq:system}) is the point $(0,0)$ i.e. there is a consensus amongst the groups. When $Kc\alpha >1$ this trivial fixed point loses its stability and we expect to see oscillations in the system corresponding to a limit cycle attractor (the trajectory cannot diverge to infinity since the function $\tanh(x)$ is bounded).

What is also interesting in this case is how the sustained oscillations change as we modulate $\alpha$ or $c$. As before we choose to modulate $c$ via the parameter $m$. 
In Fig.~\ref{fig:asymmetric_freq} we show both frequencies and amplitudes as functions of $c\alpha$ with either $\alpha$ or $m$ modulation. A supercritical Hopf bifurcation takes place at the point $Kc\alpha=1$ and the frequency of the emerging periodic orbit at the critical point should be equal to  $f_{crit}=ca\sqrt{K_{xy}K_{yx}}/2\pi\approx 0.1592$. Although the oscillations are highly non-linear (due to the $\tanh(x)$ term), the mean field predictions are showing a good qualitative match to agent-based simulations.  The frequency $f$ is slightly different in the overcritical region as compared to the critical value $f_{crit}$ that is in agreement with the theory of Hopf bifurcation \cite{ott2002chaos}. For large values of $c\alpha$ the amplitude of oscillations saturates  as the function of the parameter $\alpha$ and is a linear function of the parameter $c$. This  behaviour is similar to  plots  at Fig. \ref{fig:sym_crit} and it related to scaling observed for  the asymptotic steady state solution $x_{t \to \infty} $, see Eq.~(\ref{eq:xapprox2}).                             

While this scenario might be slightly less obvious to interpret we do believe there are certain parallels to be drawn here. It may seem as though one groups is a \textit{trend setter} while the other are \textit{followers}. In such a case there is a very similar sort of a feedback dynamic that we observe in our model. One group - the followers - is positively oriented towards the other - the trend setters - as they look up to them and would like to be, act, think like them etc. On the other hand, the setters share a negative attitude towards the followers in this context. While they might appreciate the following they would very much want to move away from it in terms of the opinion in question. This leads to this chasing and oscillating behaviour. However, should the attitudes magnitudes \textit{within} the groups be not strong enough the dynamic simply dies down as neither the followers are interested in following nor the setters in trend setting.

\section{Conclusions}
In this paper we consider a temporal bi-layer echo chamber and polarisation model on complex networks inspired by the mono-layer model introduced by Baumann et al. We recognise that there is both a precedent and apparent value in studying scenarios where two clearly cut groups - or layers in a network -  are interacting with one another. Understanding how layered complex networks evolve in various environments in context of opinion dynamics can help us better prepare for studying in detail such prominent real-world social phenomena as misinformation campaigns or echo chambers.

We formulate the dynamics equations for the bi-layer system (\ref{eq:2groups}) and then provide a mean field analysis that uncovers interesting possible scenarios. The nature of system's behaviour is different depending on the coupling between the layers. We categorise those coupling as symmetric and non-symmetric with a special case of an added external bias also considered. In more detail there is a negative symmetric case where the groups do not like each other, opinion oscillations where one group likes the other, however, the feeling is not mutual, the aforementioned external bias where we consider the other group as an external bias acting upon a mono-layer system and finally a weak positive coupling where there is an attraction between the groups, however, not as strong as withing them.

When the two layers are weakly yet positively coupled we see that there exists a critical value of the coupling parameter that causes the system to experience a sudden shift in the opinions. In this case we observe that there is a transition from a polarised state to a one side consensus (or a radicalised state) where all agents (from both layers) share similar and non-zero opinion. Similarly to the previous case the match between the mean field theory and simulations is qualitatively satisfying, however, for larger values of the control parameter the predictions as to when the transition should happen diverge from the results of numerical experiments. 

In the opposite polarisation scenario, i.e., negative symmetric, we observe that a coexistence of two groups with different (opposite) opinions is possible. The system undergoes a phase transition from a neutral consensus - where the two layers' opinions merge at zero - to a polarised state - where the two groups coexist each of them having their own opinion, opposite to the other groups. The details of this pitchfork bifurcation and the asymptotic behaviour of the system depend on whether we modulate the non-linearity parameter $\alpha$ or the combined social influence parameter $c$, or the coupling parameters $K_{xy}, K_{yx}$, however, in both cases the mean field approximation gives us a satisfying fit to agent-based simulations.

In the case of a single layer with an external bias present we postulate that it might be possible to model either a background of some sort or the second layer for that matter as simply a cumulative effect in the form of such an external bias. We find that the behaviour here is in a not very dissimilar fashion to the weak positive coupling scenario. Namely there exists a critical value of said bias that when the system is subject to it a sudden change to an opposite opinion is possible and the cusp catastrophe is apparent. For small values of the control parameter we find a decent match of mean field approach and agent-based simulations, however, for larger values the two diverge in the prediction as to when the transition should occur, most likely due to the finite size of the simulated system.

Finally, when the coupling parameters are set anti-symmetrically, in the sense that one is positive and one negative, we detect a transition from dampened to sustained oscillations of the layers' opinions - a supercritical Hopf bifurcation.
In a way one might say that one group is "chasing" the other with their opinions, while the other is trying to get away. We additionally find that the oscillations are highly non-linear as the frequency \textit{decreases} with control parameter as opposed to \textit{increasing} as one would expect from a linear oscillator. At the same time the amplitude rises with the control parameter. We believe the amplitude here plays the role of a sort of barrier for the system to overcome and so the higher the barrier the longer it takes to be overtaken thus the frequency of the oscillations increase.

With each scenario we have drawn parallels to real world to illustrate what these results could mean for understanding the dynamics of our societies. We understand that there are limitations with both the model and the approach in general as it can be often difficult to construct reproducible experiments in sociological context, however, we firmly believe that seeing where certain assumptions can lead us is an important and crucial building block of science.

\begin{acknowledgments}
This research was supported by an IDUB against COVID-19 project granted by Warsaw University of Technology under the program Excellence Initiative: Research University (IDUB). JAH wase partially supported by the Russian Science Foundation, Agreement No. 17-71-30029 with co-financing of the Bank Saint Petersburg.
\end{acknowledgments}


\bibliography{apssamp}

\end{document}